\documentclass[a4paper,11pt]{article}	

\usepackage{classeJBArxive}

\usepackage{arydshln}
\usepackage{algorithm}
\usepackage{algorithmic}

\usepackage{nomencl}% Nomenclature package
\makenomenclature
\renewcommand*\nompreamble{\begin{multicols}{2}}
\renewcommand*\nompostamble{\end{multicols}}

\newcommand{\degC}{^{\,\circ}C}

\title{
\vspace{-1.5cm}
Design and Experimental Validation of an Urban Microclimate Tool Integrating Indoor-Outdoor Detailed Longwave Radiative Fluxes at District Scale
\vspace{4pt}
}

\author{Marie-Hélène Azam{a}$^{\ast}$, Julien Berger\textsuperscript{b}, Edouard Walther\textsuperscript{a}, Sihem Guernouti\textsuperscript{c}\\
\date{\today\vspace{-0.5cm}}}

\begin{document}
%\frontmatter

\maketitle

\begin{center}
\small
\textsuperscript{a} ICube Laboratory UMR 7357, Team GCE, Strasbourg University, INSA Strasbourg, CNRS, 67000, Strasbourg, France \\
\textsuperscript{b} Laboratoire des Sciences de l’Ingénieur pour l’Environnement (LaSIE), UMR 7356 CNRS, La Rochelle Université, CNRS, 17000, La Rochelle, France\\
\textsuperscript{c} CEREMA BPE, Nantes F-44262, France\\
$^{\ast}$corresponding author, e-mail address : marie-helene.azam@insa-strasbourg.fr\\
\end{center}

\begin{abstract}

Numerical simulation is a powerful tool for assessing the causes of an Urban Heat Island (UHI) effect or quantifying the impact of mitigation solutions on outdoor and indoor thermal comfort. For that purpose, several models have been developed at the district scale. At this scale, the outside surface energy budget is detailed, however  building models are very simplified and considered as a boundary condition of the district scale model. This shortcoming inhibits the opportunity to investigate the effect of urban microclimate on the inside building conditions. The aim of this work is to improve the representation of the physical phenomena involved in the building models of a district model. For that purpose, the model integrates inside and outside fully detailed long-wave radiative flux. The numerical model is based on finite differences to solve conduction through all the surfaces and the radiosity method to solve long-wave radiative heat fluxes inside and outside. Calculated temperatures and heat fluxes are evaluated with respect to \textit{in situ} measurements from an experimental demonstrator over 14 sensors and a 24-day period. Results are also compared to state-of-the-art models  simulation tool show improvement of the RMSE of $0.9 \ \mathsf{^{\,\circ}C}$ to $2.1 \ \mathsf{^{\,\circ}C}$ on the surface temperature modeled. 


\end{abstract}

\section{Introduction}
\label{S1:Intro}

With global warming, heat waves are on the increase. This phenomenon is amplified in cities with dense mineral environments by the urban heat island (UHI) effect. It has consequences not only on the comfort of the users, but also on building inside conditions (air temperature) and energy consumption (air conditioning) in summer. Cities must adapt to these local climate conditions by proposing mitigation solutions in their development scenarios, and buildings must be designed considering them. 

Numerical simulation is then a powerful tool to understand the consequences of a development scenario and compare the impact of different solutions under consideration on the outside comfort or on the building inside conditions. For that purpose, several models have been developed, at the district scale, as \texttt{SOLENE-Microclimat} \cite{musy2015use}, \texttt{Envi-met} \cite{yang2013evaluation} or \texttt{LASER/F} \cite{kastendeuch2017thermo}. On one hand, the modeling of the urban energy balance is very detailed considering short and long wave (LW) radiative heat balance, isotherm fluids dynamics computation, and heat conduction through the solid elements of the building wall and ground soils. On the other hand, building models are very simplified considering mono-zone air model (or fixed inside temperature), long wave exchanges linearization, thermal bridges or systems are not considered, etc. They are considered as a boundary condition of the district scale model. This limits the opportunity to study the effect of urban microclimate on the inside building conditions. 

The aim of this work is to improve the representation of the physical phenomena involved in the building model of a district model by proposing a mathematical model more detailed than ones from the state-of-the-art. The physical problem considers : \emph{(i)} outside short wave radiative heat flux \emph{(ii)} outside long wave radiative heat flux for the whole urban scene and the sky \emph{(iii)} 1D conduction through all the elementary surfaces of the urban scene \emph{(iv)} inside building zone long-wave radiative heat fluxes \emph{(v)} inside building zone air temperature calculation.

Before using those types of models to evaluate the performance of mitigation technologies, their reliability needs to be verified \cite{asme_vv_asme_2009}. First, the verification of the building model is assessed on a theoretical case with a reference solution. It ensures that the mathematical model computes correctly the solution of the problem. The second study aims at validating the model by comparing the numerical prediction with experimental observations obtained from a reduced-scale experimental demonstrator presented in \cite{djedjig2013experimental}. It guarantees that the physical phenomena are correctly represented. Note that a comparison with a state-of-the-art model is also proposed to evaluate the increase of the model reliability.

After a presentation of the physical problem and the mathematical models associated in Section \ref{sec:Physical_problem}, 
section \ref{sec:theoretical_case_study} presents the verification case study. The models are then applied to a realistic case study, in section \ref{sec:Practical application}. Results are compared to experimental data and state-of-the-art model \texttt{SOLENE-microclimat} simulation tool as a reference tool in the field. Finally, a discussion is proposed on the model performances and opportunities. 

\section{Methods}
\label{sec:Physical_problem}

First, the physical problem and the classical architecture of a micro-climatic simulation tool are recalled. It usually consists of five mathematical models : an enclosure model, an inside building air temperature model, an outdoor airflow model, a long-wave and a short-wave radiative model. Every model will be integrated in the following work except for the outdoor airflow model. For the latter, a simplified approach is used considering a surface exchange coefficient calculated from the weather data wind speed. 

\begin{figure}
    \centering
    \includegraphics[width=0.8\textwidth]{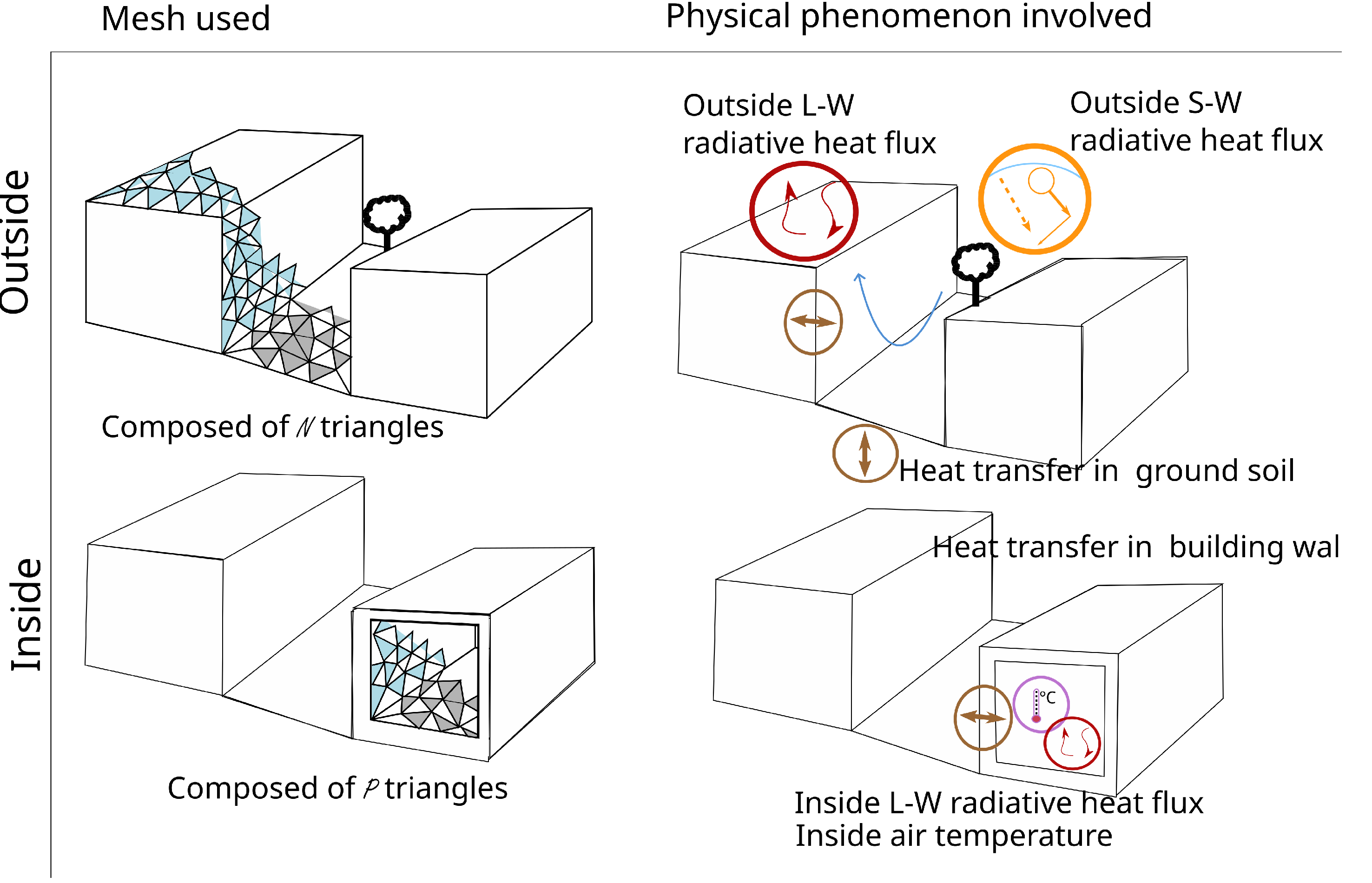}
    \caption{Schematic representation of the different mathematical models of a microclimat simulation tool at district scale and the mesh.}
    \label{fig:modeles_math_mesh}
\end{figure}

Each mathematical model then relies on a resolution method to achieve a numerical model. For the first mathematical models (i.e. enclosure model, inside building air temperature model), a system of partial differential equations is obtained and solved using the finite difference method with an implicit scheme. For the radiative models, the radiosity method is used to solve the problem \cite{incropera2022fundamentals}. 

The simulation tool is therefore composed of the above-mentioned models are coupled together. For that purpose, the urban scene is composed of the enclosures (building walls, roof, outside ground soils) and inside building zones. The geometry is composed of the surfaces meshed as presented in Figure \ref{fig:modeles_math_mesh}. The mesh is composed of  $\mathcal{N}$ triangles for the enclosures, and each inside building zone is divided into $\mathcal{P}$ triangles. For each piece of the total mesh, the thermo-radiative balance is derived and one of the above-mentioned models has been implemented. To illustrate this process, the next Sections describe further the mathematical models and the link between them.
 
\subsection{Heat transfer in the building wall}

Building walls are part of the enclosures of the urban scene. The physical problem considers one-dimensional transient heat conduction equation for a time interval $\Omega_{\,t} \,  \ = \  \, \bigl[\, 0 \,,\, \tau \, \bigr]$ and space interval $\Omega_{\,w } \,  \ = \  \, \bigl[\, 0 \,,\, e_w \, \bigr]$, where $e_w$ designs the wall thickness and $\tau$ the time horizon. For each triangles related to the walls, the physical problem can be formulated as:
\begin{subequations}
\label{eq:heat_transfert_wall}
\begin{align}
    c_w \, \frac{\partial T }{\partial t}  & \ = \  \frac{\partial}{\partial  x} \, \biggl(\, k_w \, \frac{\partial T}{\partial x} \, \biggr) \, && \\[4pt]
    - \, k_w \ \frac{\partial T }{\partial x}   & \ = \  q_{\, out} (t)  \ - \   h_{\, out}\bigl(\,v(\,t\,)\,) \, \Bigl(\, T  \ - \ T_{\,out} (t) \, \Bigr) \,,  && x \ = \  0 \,, \\[4pt]
    k_w  \ \frac{\partial T}{\partial x}   & \ = \ q_{\, in} (t) \ - \ h_{\, in} \, \Bigl(\, T  \ - \  T_{\,in}(t) \, \Bigr) \,,   && x \ = \  e_w \,.\\[4pt]
    T  & \ = \  T_{\,0}  \,,  && t  \ = \ 0 \,.
\end{align}
\end{subequations}
where 
$T$ is the temperature that depends on spatial $x$ and time $t$ variables, 
$k_w$ is the thermal conductivity in the wall that depends on spatial $x$ variable, 
$c_w$ is the volumetric heat capacity in the wall that also depends on spatial variable, and 
$T_{\,0}$ is the initial temperature profile. 
The inside air temperature $T_{\,in} (t)$ is either computed by Eq.~\eqref{eq:inside_energy_balance} and the model described in Section~\ref{sec:inside_air_temperature} or imposed. The outside air temperature $T_{\,out} (t)$ is given by the climatic data. The convective heat transfer coefficient is noted $h_{\, out}$ for the outside surface and $h_{\, in}$ for the inside surface. Those coefficients can be imposed or calculated thanks to correlations. They vary for each triangle of the mesh and time. 
Inside and outside, the net radiative heat flux balance respectively noted $q_{in} (t)$ and $q_{out} (t)$ involves short-wave radiative heat flux density noted $q_{SW} (t) $ ; inside and outside long-wave radiative heat flux densities noted $q_{LW, in} (t) $ and $q_{LW, out} (t) $. Those four heat flux densities are calculated for each triangle of the mesh and are time dependent. 
\begin{align}
    q_{in} (t) & =  q_{\, LW, \ in} (t) \\[4pt]
    q_{out} (t) & = q_{SW} (t) +q_{\, LW , \ out } (t)
\end{align}
The conductive heat flux density in the wall is defined as: 
\begin{align}
    - \, k_w \ \frac{\partial T }{\partial x}   & \ = \  q_{\, c} (t) && x \ = \  0 \,,
\end{align}
The problem is rewritten in a dimensionless form before applying a resolution method. Those equations are presented in Appendix~\ref{Appendix1}. 

\subsection{Heat transfer in the ground soil}

Ground soils are part of the enclosure of the urban scene and the building inside ground floor. The physical problem considers one-dimensional transient heat conduction equation for a time interval $\Omega_{\,t} \,  \ = \  \, \bigl[\, 0 \,,\, \tau \, \bigr]$ and space interval $\Omega_{\,s} \,  \ = \  \, \bigl[\, 0 \,,\, e_s \, \bigr]$, where $e_s$ designs the soil column depth studied and $\tau$ the time horizon. For each triangles related to the ground soils, the physical problem can be formulated as:
\begin{subequations}
\label{eq:heat_transfert_soil}
\begin{align}
    c_s \, \frac{\partial T }{\partial t}  & \ = \  \frac{\partial}{\partial  x} \, \biggl(\, k_s \, \frac{\partial T}{\partial x} \, \biggr) \, && \\[4pt]
    - \, k_s \ \frac{\partial T }{\partial x}   & \ = \  q_{out} (t)  \ - \   h_{ \, out}\bigl(\,v(\,t\,)\,) \, \Bigl(\, T \ - \ T_{out} (t) \, \Bigr) \,,  && x \ = \  0 \,, \\[4pt]
   T ( \ x \ = \ L \ )  &  \ = \ T_{\infty},   && x \ = \  e_s \,,\\[4pt]
    T  & \ = \  T_{\,0}  \,,  && t  \ = \ 0 \,.
\end{align}
\end{subequations}
where 
$k_s $ is the thermal conductivity in the ground soil column that depends on spatial $x$ variable, 
$c_s $ is the volumetric heat capacity in the ground soil that also depends on spatial variable. 
The problem is rewritten in a dimensionless form before applying a resolution method. Those equations are presented in \ref{Appendix1}.

\subsection{Inside air temperature}
\label{sec:inside_air_temperature}

In order to get a complete building inside energy balance, the previous building wall model and ground soil model are coupled to an inside air temperature model.  The air model will provide an energy balance over the inside air volume of one zone of the building. It can be described by the following equation:
\begin{align}
\label{eq:inside_energy_balance}
    c_{air} \, V_{air} \, \frac{ \mathrm{d}  T_{ \, in} (t) }{ \mathrm{d}  \, t} & \ = \ q_v \ + \sum_{j \ = \ 1}^{\mathcal{P}} h_{ \,in} \, S_{\,j} \, \Bigl(\, T_{j}  \ - \ T_{\,in} (t) \, \Bigr), 
\end{align}
 with $c_{air}$ the volumetric heat air capacity , $V_{air}$ the zone air volume, $S_{\,j}$ the inside air temperature and $q_v$ a source or sink term that represents the heat generation or loss induced by HVAC systems, infiltrations or occupants. As illustrated in figure \ref{fig:modeles_math_mesh} , the quantity $\mathcal{P}$ designed each triangle of the inside mesh air zone.  The problem is rewritten in a dimensionless form before applying a resolution method. Those equations are presented in \ref{Appendix1}.

\subsection{Short-wave radiative heat fluxes}

Incident solar radiations are calculated by the SOLENE \cite{miguet2002daylight} radiative simulation software, taking into account direct and diffuse radiation emitted with inter-reflections in the urban scene.  To calculate the inter-reflections, the radiosity method is used. It considers only totally diffuse and isotropic reflections. 

\subsection{Long-wave radiative heat fluxes}

Long-wave radiative flux depends on the temperatures of neighboring surfaces.  \cite{noilhan1981model} gives a description of the long-wave radiation equation for an outside surface. We complete this description by integrating the radiosity method into the equations.

\subsubsection{Inside long-wave radiative heat fluxes}
For one building zone, the model includes calculation of the inside long-wave radiative balance calculating the following sum for each triangle $i$ of the zone mesh (containing a total of $\mathcal{P}$ triangles)  :
\begin{equation}
    q_{LW,\,  in, i}= \sigma   \sum_{j=1}^{\mathcal{P}}  F_{i, \, j} \left(  \epsilon_j T_{j}^{4} -   \epsilon_i T_{i}^{4} \right)
    \label{Eq:GLO_surf}
\end{equation}
where $\sigma \  = \  5.67 \times 10^{-8} \mathsf{W}. \mathsf{m}^{-2}.\mathsf{K}^{-4} $ is the Stefan Boltzmann constant, $F_{i, j }$ is the form factor between triangle $i$ and $j$, and $\epsilon_j$ and $\epsilon_i$ are the emissivity of the triangle $i$ and $j$. 
For that purpose, the radiosity method is used \cite{incropera2022fundamentals}. The radiosity of the triangle $i$ noted $J_i$ can be defined by the energy radiated by the surface, including the energy emitted by the surface itself and the energy reflected by incident radiation. 
\begin{equation*}
    J_i = \epsilon_i \cdot \sigma \cdot T_i^4 + (1- \epsilon_i) \cdot H_i
\end{equation*}
where $H_i$ is the irradiation of the triangle $i$ including contributions from other surface radiosities, multiply by the form factors $F_{j,i}\,$: 
\begin{equation*}
    H_i = \sum_{j \ = \ 1 }^{\mathcal{P}}  F_{i, \, j} \cdot  J_j
\end{equation*}
For all the discretized surfaces $\mathcal{P}$ of the inside studied zone, the following system of equations relating  the radiosity of each triangle needs to be solved : 
\begin{equation}
\label{eq:LW_in}
    J_i \ - \ (1 - \epsilon_i ) \sum_{j \ = \ 1 }^{\mathcal{P}}  F_{j, \, i} \cdot J_j  \ =  \ \epsilon_i \ \cdot \ \sigma \cdot T_i^4 , \ \qquad \forall \, i \in \bigl\{\,1\,,\,\ldots\,,\,\mathcal{P} \,\bigr\}
\end{equation}
Once the previous equation system is solved, the long-wave net radiative flux density exchanged by a triangle $i$ can be calculated as: 
\begin{equation}
\label{eq:LW_inside}
    q_{LW,\,  in, i}= J_i \ - \ H_i
\end{equation}
The form factors are calculated with the \texttt{pyviewfactor} python library \cite{bogdan2022calcul}. Given the iterative nature of the heat balance, it is difficult to take into account all infrared inter-reflections, as in the case for short-wave radiation. Therefore, the model is limited to a single reflection, which is considered satisfactory by most models developed in the literature \cite{masson2000physically, musy2015use} given that the emissivities of the urban surfaces considered are generally around $0.9$. 

\subsubsection{Outside long-wave radiative heat fluxes}
For the urban scene, the model includes the outside long-wave radiative balance with two terms: the contribution of the urban scene and the contribution of the sky.  The outside irradiation of the surface $i$ is the sum of the contributions of radiosities from other surfaces and from the sky:
\begin{equation}
    H_i \ = \ \sum_{j \ = \ 1 }^{\mathcal{N}} J_j \cdot F_{j, \, i} \ +  \ F_{i, sky} \ \cdot \ J_{sky}
\end{equation}
For all the discretized surfaces $\mathcal{N}$ of the urban scene, the following linear system of equations that links the radiosity of each surface and the sky is obtained: 
\begin{equation}
\label{eq:LW_out}
    J_i \ - \ (1 - \epsilon_i ) \sum_{j \ = \ 1 }^{\mathcal{N}} J_j \cdot  F_{j, \, i} \ - \ (1 - \epsilon_i ) \cdot  F_{i, sky} \ \cdot \ J_{sky} =  \ \epsilon_i \ \cdot \ \sigma \cdot T_i^4 , \qquad \forall \, i \in \bigl\{\,1\,,\,\ldots\,,\,N \,\bigr\}
\end{equation}
The form factor between the surface and the sky noted $F_{i, sky}$ is calculated as the complementary part of the sum of the surface form factors:
\begin{equation*}
    F_{i, sky} = 1 - \sum_{j \ = \ 1 }^{\mathcal{N}} F_{i, \, j}
\end{equation*}
The sky radiosity can either be measured or calculated given the sky temperature as : $J_{sky} \ = \ \sigma \ . \ T_{sky}^4$. Once the previous equation system is solved, the long-wave net radiative flux density exchanged by a triangle $i$ can be calculated as: 
\begin{equation}
\label{eq:LW_outside}
    q_{LW,\,  out, \, i}= J_i \ - \ H_i
\end{equation}

\subsection{Algorithm implementation}

After meshing the urban scene as presented in Figure \ref{fig:modeles_math_mesh} for each piece of mesh, the thermo-radiative balance is derived using previously described equations. For that purpose, at each time step $t^{\,n}$, all models exchange parameters like surface temperature and net radiative heat flux balance with the other models. During the interval $t \in [t^{\,n},t^{n\ + \  1}]$, each model computes the field of interest, which consists of the temperature profiles. 

Once the mathematical model has been defined and rewritten into a dimensionless form, a numerical method is chosen to solve the problem. To achieve this, a 1D mesh is defined behind each triangle of the surface 2D mesh and the differential operators link the spatial and temporal meshes. The definition of the temporal operator has an impact on the strategy chosen to couple the two models. 
With the implicit scheme, the solution depends on the boundary conditions of the following time step $t^{n+1}$. A solver function is defined to solve the previous equations noted  $\mathcal{F}_w$ for the building walls Eq.~\eqref{eq:heat_transfert_wall}, $\mathcal{F}_s$ for the ground soil Eq.~\eqref{eq:heat_transfert_soil}, $\mathcal{F}_{air}$ for the inside air temperature Eq.~\eqref{eq:inside_energy_balance}, $\mathcal{F}_{LW, in}$ for the inside long-wave radiative heat fluxes Eq.~\eqref{eq:LW_inside},  $\mathcal{F}_{LW, out}$ for the outside long-wave radiative heat fluxes Eq.~\eqref{eq:LW_outside}.
\begin{subequations}
\begin{align}
 \text{Wall solver:} \quad   T^{\ n+1 } \ & = \ \mathcal{F}_{w} ( \, x, T^{\ n  }, \, T^{\ n+1 }_{ \ out }, \, T^{\ n+1 }_{\ in }, \, q^{\ n+1 }_{\ out }, \, q^{\ n+1 }_{\ in }) ,\\[4pt]
 \text{Ground solver:}  \quad  T^{\ n+1 } \ & = \ \mathcal{F}_{s} ( \, x, T^{\ n  }, \, T^{\ n+1 }_{ \ out }, T_{\infty}^{n+1}\, q^{\ n+1 }_{\ out }) ,\\[4pt]
 \text{Inside air solver:}  \quad  T_{air}^{\ n+1 } \ & = \ \mathcal{F}_{air} ( \, x,  T^{\ n+1 }, \, T^{\ n }_{\, in }) ,\\[4pt]
  \text{Inside radiation solver:} \quad   q_{LW, out}^{\ n+1 } \ & = \ \mathcal{F}_{LW, out} ( \, F,  T(x=0)^{\ n+1 }) ,\\[4pt]
  \text{Outside radiation solver:} \quad   q_{LW, in}^{\ n+1 } \ & = \ \mathcal{F}_{LW, in} ( \, F,  T(x=e_w)^{\ n+1 }) \,.
\end{align}
\end{subequations}
The boundary conditions then involve the variables at the resolution time step ($t^{n+1}$). Since the radiative flux $q_{LW}$ depends on surface temperatures at the resolution time step ($t^{n+1}$), the resulting system of equations is non-linear. The resolution requires the use of two fixed-point algorithms: (1) One for the outside urban scene thermo-radiative budget. The iterative process that successively solves : the all enclosure (ground soils and walls) and the outside radiative heat fluxes. The convergence of the algorithm is verified for each surface temperature of the urban scene.  (2) A second loop is necessary for the building studied zone, the model iterates to solve the walls, the inside long-wave radiative fluxes and air temperature until it converges. This method is also known as strong coupling \citep{hensen1995modelling}. 

% j'ai essayé de garder que les étapes essentielles pour faire aparaitre les deux boucles itératives.
\begin{algorithm}
\caption{Fixed-point algorithm to calculate ($T^{n+1}, T_{in}^{n+1}$)}
\begin{algorithmic}
%\STATE $T^{k}=T^{n+1}$
%\STATE $T_{air, in}^{k}=T_{air, in}^{n+1}$
%First iterative loop:
\WHILE{ $|| T(x=0)^{k+1}-T(x=0)^{k}  || < \eta$  }
%Initialization of the variables from previous iteration:
%\STATE  $q^{\ k }_{GLO, \ out } = q^{\ k +1 }_{GLO, \ out }$
%\STATE $T^{k}=T^{k+1}$
  \STATE      $T^{\ n+1 } \  = \ \mathcal{F}_{w} ( \, x, T^{\ n  }, \, T^{\ n+1 }_{ \ out }, \, T^{\ n+1 }_{\ in }, \, q^{\ n+1 }_{\ out }, \, q^{\ n+1 }_{\ in }) $
   \STATE     $T^{\ n+1 } \  = \ \mathcal{F}_{s} ( \, x, T^{\ n  }, \, T^{\ n+1 }_{ \ out }, T_{\infty}^{n+1}\, q^{\ n+1 }_{\ out }) $
%Second iterative loop for the building studied zone:
\WHILE{  $|| T(x=e_w)^{m+1}-T(x=e_w)^{m}, T_{\,in}^{m+1}-T_{in}^{m}  || < \eta$ }
%\STATE $q^{\ m }_{GLO, \ in } = q^{\ m +1 }_{GLO, \ in }$
\STATE $T^{m+1} = \mathcal{F}_{w} (x,  \, T^{n } , \, T_{\,in}^{n +1 } , \, T_{\,out}^{n +1 }, \, q^{\ k }_{out }, \, q^{\ m }_{ in })$
\STATE $T_{in}^{m+1} = \mathcal{F}_{air}(x,  \, T^{m +1 } , \, T_{\,in}^{n })$
%\STATE $T^{m}=T^{k+1}$
\STATE $q^{\ n +1 }_{LW, \ in } =  \mathcal{F}_{LW, in}(\, F, \, T(x=e_w)^{k +1 } )  $
%\STATE $T_{\,in}^{k}= T_{\,in}^{k+1}$
\ENDWHILE
\STATE $T^{k+1}=T^{m+1}$
\STATE $q^{\ n +1 }_{LW, \ out } =  \mathcal{F}_{LW, \ out }(\,F, \  T(x=0) ^{k +1 }, \ T^{n +1 }_{sky} )  $
\ENDWHILE
%\STATE $T^{n+1}=T^{k+1}$
%\STATE $T_{air, \, in }^{n +1 }= T_{\,in}^{k+1}$
%\STATE $q^{\ n +1 }_{GLO, \ out } = q^{\ k +1 }_{GLO, \ out }$
%\STATE $q^{\ n +1 }_{GLO, \ in } = q^{\ k +1 }_{GLO, \ in }$
\end{algorithmic}
\label{algo:cosim}
\end{algorithm}

The enclosure problem, described by the equations \ref{eq:heat_transfert_wall} and \ref{eq:heat_transfert_soil}, is solved by an implicit \textsc{Euler} scheme. The spatial mesh is one-dimensional, uniform and composed of 101  nodes. The spatial mesh and the temporal mesh are characterized by the variables $\Delta x$ and $\Delta t$. The dimensionless time step is designed by $\Delta t^{\star}$. The parameter $\eta$ is set to : $\eta = 0.01 \Delta t^{\star}$ as proposed by \cite{gasparin2018improved}. 

To simplify the development, distribution and easily integrate existing tools (as \texttt{pyviewfactor} python library), the code is all developed in python.

\subsection{Assessment methodology}
\label{sec:Methods}

The overall aim of this work is to improve the representation of physical phenomena important for coupling the building external and internal environment such as long-wave radiative heat flux.  To assess the models in different situations, several case studies will be presented.  

For each step of the assessment process, the results of the model $T_{num}(x,t)$ are then compared to the results of a reference solution $T_{ref}(x,t)$ that can arise from analytical computations or experimental results. For each case study, the indicator chosen is the $ \ell_2$ norm (or Root Mean Square Error) noted $\varepsilon_2$; it is computed as a spatial or time function by the following discrete $ \ell_2$ formulation, where $N_x$ and $N_t$ are the number of space or time grid elements respectively:
\begin{subequations}
\begin{align}
        \varepsilon_{\,2}(\,t\,) \ &  = \   \sqrt{ \frac{1}{N_x} \sum _{0}^{N_x} \left [ T_{num}(x,t)-T_{ref}(x,t) \right ]^2  } ,\\[4pt]
            \varepsilon_{\,2}(\,x\,) \ &  = \  \sqrt{\frac{1}{N_t} \sum _{0}^{N_t} [ T_{num}(x,t)-T_{ref}(x,t)]^2}
\end{align}
\end{subequations}
The global error is given by the maximum of the previous functions $\varepsilon_{\,2}(\,t\,)$ and $\varepsilon_{\,2}(\,x\,) $ as described hereafter: 
\begin{equation}
\varepsilon_{\,\infty} \ = \ \max _t \bigl(\, \varepsilon_{\,2}(x) \,\bigr) 
\end{equation}

\section{Building model verification case studies }
\label{sec:theoretical_case_study}

%\section{Verification case studies }
%\label{sec:theoretical_case_study}

%\subsection{Long-wave radiative fluxes validation }
%This verification case study aims to verify the long-wave radiative model defined by the equations  \ref{eq:LW_out} developed in the micro-climatic simulation tool. 

%\mahe{Présentation d'un cas d'étude de référence pour valider le GLO à l'extérieur. Proposition d'Edouard de résoudre un exercice de Incopera (ouvrage de référence en rayonnement). Travail en cours du côté d'Edoaurd.  }
 
%\subsubsection{Case study description }
%\mahe{ à compléter Edouard :) }

%\subsubsection{Results}
%\mahe{ à compléter Edouard :) }

%\subsection{Building model verification }

This verification case study aims to verify the building model resolution defined by the equations  \ref{eq:heat_transfert_wall} and  \ref{eq:inside_energy_balance} developed in the micro-climatic simulation tool. 

\subsection{Description}
%\subsubsection{Description}

\begin{figure}[htp]
    \begin{minipage}[c]{.46\textwidth}
  \centering
    
    \subfigure[Boundary conditions applied for the theoretical case study]{
    \includegraphics[width=1\linewidth]{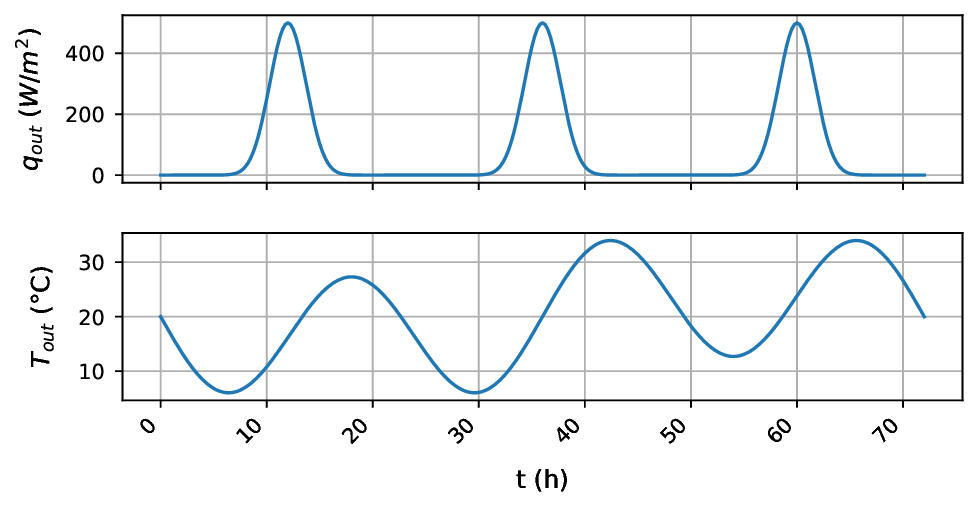} \label{fig:u_o_i_qe}
  }
    \end{minipage}
    \hfill%
    \begin{minipage}[c]{.46\textwidth}
    
  \subfigure[Error as a function of the time step for two mesh]{
  \includegraphics[width=1\textwidth]{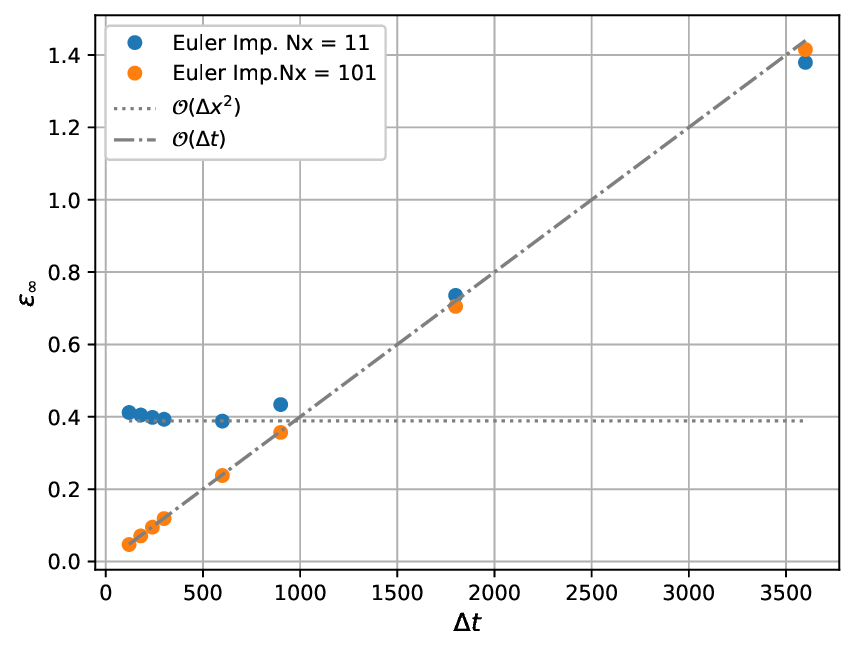} \label{fig:Err_dt_dx}
  }
        \end{minipage}
  \caption{Case study set up : boundary conditions, and mesh selection criterion for the model.}
  %\label{fig:Comp_mur_int_ext}
\end{figure}

The theoretical case study is inspired from \cite{gasparin2018improved}. It assumes a cubic building, composed of 4 walls in contact with the outside environment. The walls (dimensions : $ 3.00 \ \times \ 4.00 \ \mathsf{m}$  ) are composed of a single layer of concrete of thickness $e_w = 0.20 \, \mathsf{m}$, with thermal conductivity $k_w = 1.75 \, \mathsf{W}. \mathsf{m}^{-1}. \mathsf{K}^{-1}$ and a volumetric heat capacity $c_w \ = \ 2.2 \, 10^{\,6} \, \mathsf{J}. \mathsf{m}^{-3}. \mathsf{K}^{-1}$.  The air volume (dimensions : $ 3.00 \ \times \ 4.00 \ \times \ 3.00 \ \mathsf{m}$ ) has a volumetric heat capacity $c_{air} \ = \ 1.2 \, 10^{\,3} \, \mathsf{J}. \mathsf{m}^{- 3}. \mathsf{K}^{-1}$ The initial temperature is set at $T_0 = 20 \ \mathsf{^\circ C}$. On the outside of the walls, sinusoidal variations in air temperature and net radiant flux are considered and defined as follows: 
\begin{subequations}
\begin{align}
    T_{ \,out} (t) & \ = \ T_{\, o, \, m} + \delta_{\, o,\, 1} \, \sin(\, 2 \ \pi \ \omega_{\, o,\, 1} \ t) + \delta_{\, o,\, 2} \, \sin(\, 2 \ \pi \ \omega_{\, o,\, 2} \ t) \\[4pt]
    q_{out} (t)  & \ = \ q_{ \, m} \,  \sin(\, 2 \ \pi \ \omega_{\, q,} \ t)^{\, 20}
\end{align}
\end{subequations}
The following numerical values are taken into account for external boundary conditions: 
\begin{align*}
    T_{\, o, \, m} & = \ 20 \, \mathsf{^\circ C} \,,\quad
    \delta_{\, o,\, 1} \ = \ - 4,4 \, \mathsf{^\circ C} \,,\quad
    \omega_{\, o,\, 1}  \ = \ \frac{1}{ 72 }\, \mathsf{h}^{-1} \,,\quad
    \delta_{\, o,\, 2}  \ = \ - 11,7 \, \mathsf{^\circ C} \,, \\[4pt]
    & \omega_{\, o,\, 2} \ = \ \frac{1}{ 24 } \, \mathsf{h}^{-1} \,,\quad
    q_{ \, m} \ = \ 500 \, \mathsf{W .m^{- 2}} \,,\quad
    \omega_{\, q,} \ = \ \frac{1}{ 48 } \, \mathsf{h}^{-1} \,.
\end{align*}
The convection coefficients are set to : $h_{ in} = 8.7 \, \mathsf{W}. \mathsf{m}^{- 2}. K^{-1}$ and $h_{ \, out} \ = \ 23,2 \, \mathsf{W}. \mathsf{m}^{- 2}. \mathsf{K}^{-1}$. The net external radiative flux is applied to all walls, whatever their orientation in this case study. The radiative flux balance is neglected on the inner faces of the walls. The boundary conditions assumed are shown in Figure \ref{fig:u_o_i_qe}.

For this case study, the accuracy of the model is assessed by comparing the results with a reference solution. This is generated with the \texttt{Matlab\texttrademark} open source toolbox \texttt{Chebfun}. The solution is projected on a mesh of $101$ points with a time step of $15 \ \mathsf{min}\,$ for a time horizon of $ \Omega_t = 3 \mathsf{days}$. This is a configuration similar to the one studied later in the validation case study. 

\subsection{Results}

Figure \ref{fig:Err_dt_dx} shows the evolution of the error $\epsilon_\infty$ as a function of the time step for two different meshes on one of the walls. The errors observed for the finite-difference model are in line with theory. The scheme used is accurate to order 1 for the temporal mesh and to order 2 for the spatial mesh. Thus, when the time step is small and the mesh coarse, the error is proportional to $\Delta x^2$. However, with a fine mesh, the error due to the time step becomes predominant. In this case, the error curves are proportional to $\Delta t$. This result demonstrates the importance of using the smallest possible time steps, when a fine mesh is used. The vast majority of thermal simulation tools use hourly calculation time steps, mainly due to the availability of boundary conditions (weather data), which are often given on an hourly basis. However, for certain configurations and the study of certain phenomena (including solar radiation), a sub-hourly time step is relevant to improve the representation of phenomena.

\begin{figure}[htp]
    \begin{minipage}[c]{.46\textwidth}
  \centering
  \subfigure[Outside surface temperature of the wall]{
  \includegraphics[width=1\textwidth]{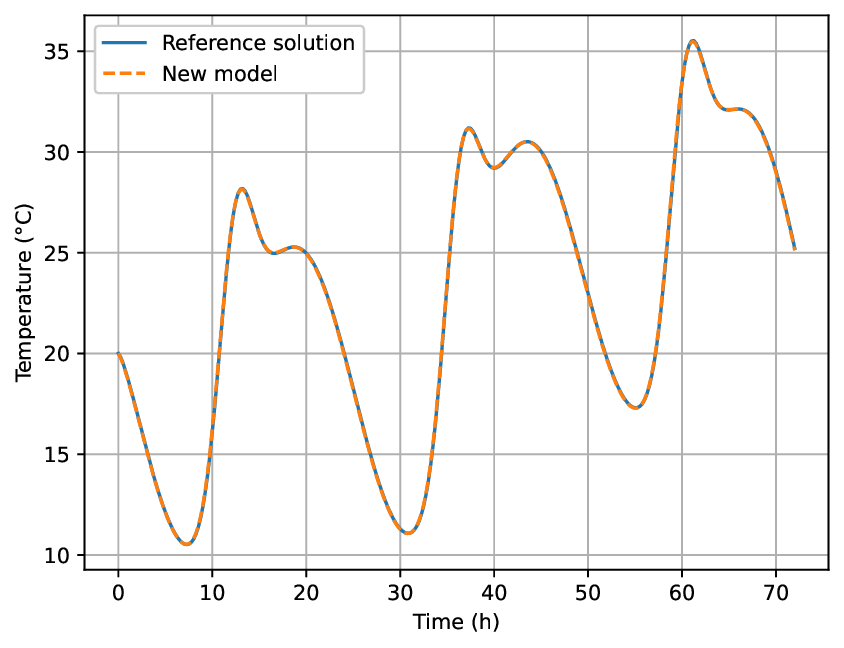}
  }
    \end{minipage}
    \hfill%
    \begin{minipage}[c]{.46\textwidth}
  \subfigure[Inside surface temperature of the wall ]{
  \includegraphics[width=1\textwidth]{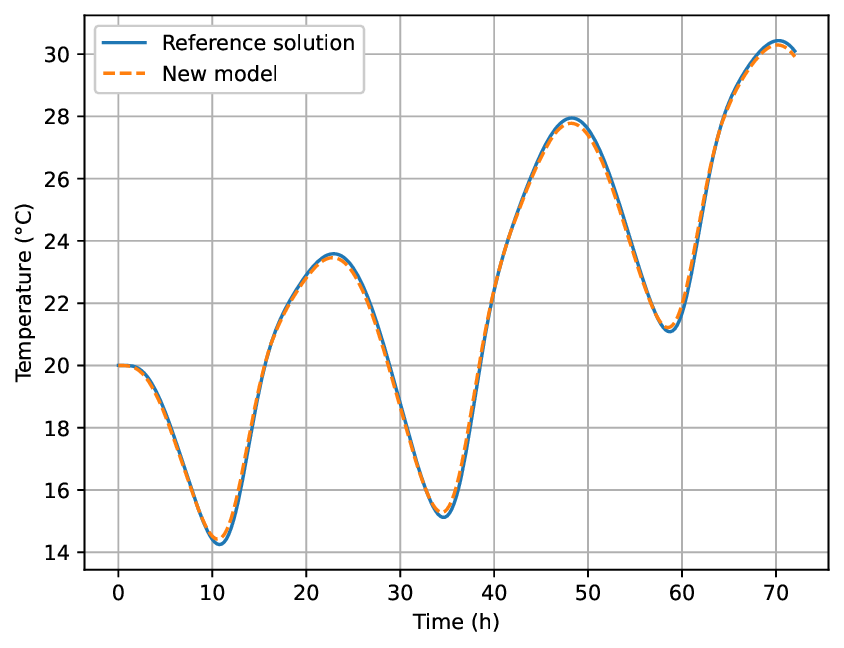}
  }
        \end{minipage}
  \caption{Comparison of surface temperatures calculated by the model and the reference solution  }
  \label{fig:Comp_mur_int_ext}
\end{figure}

Figure \ref{fig:Comp_mur_int_ext} shows the evolution of the internal and external surface temperatures of the wall. The model faithfully reproduces the reference solution. Stress dynamics are well reproduced.

\begin{figure}[htp]
    \begin{minipage}[c]{.46\textwidth}
  \centering
  \subfigure[Comparison of the air temperature calculated by the model and the reference solution]{
  \label{fig:Comp_air}
  \includegraphics[width=1\textwidth]{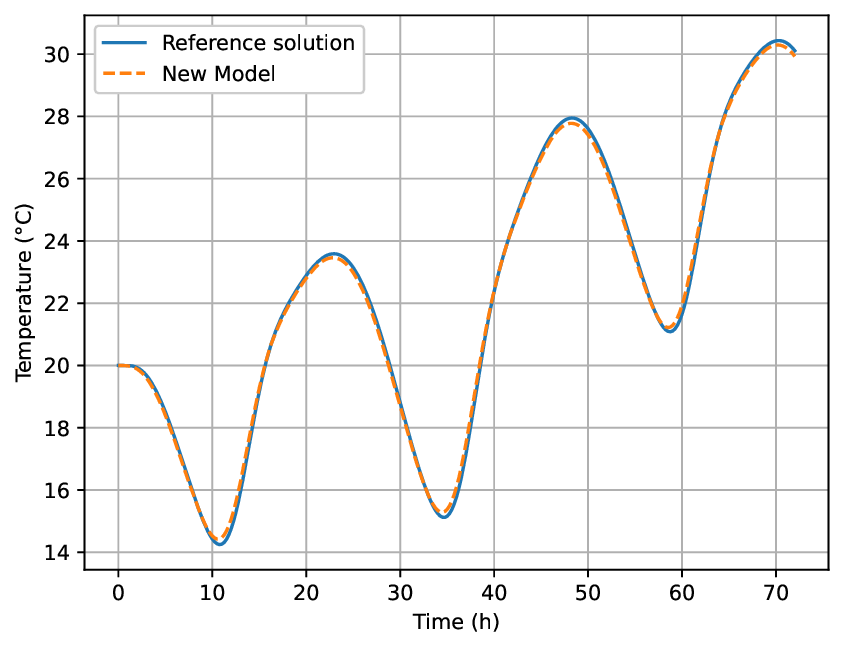}
  }
    \end{minipage}
    \hfill%
    \begin{minipage}[c]{.46\textwidth}
  \subfigure[Comparison of temperatures calculated inside the wall by the model and the reference solution at $t^n = 10 \ h $ (up) and $t^n = 27 \ h $ (down)]{
  \label{fig:Comp_mur_profils}
  \includegraphics[width=1\textwidth]{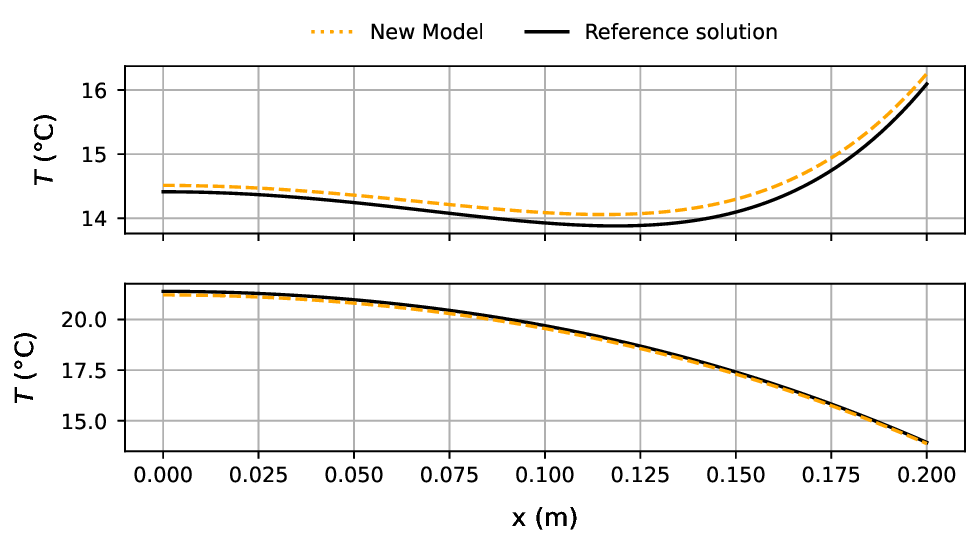}
  }
        \end{minipage}
  \caption{Comparison of air temperature and inside temperature profile calculated by the model and the reference solution  }
  %\label{fig:BC_Case_study}
\end{figure}

Figure \ref{fig:Comp_mur_profils} shows the temperature profiles inside of the wall for two distinct time steps: $t^{n}=10h$ and $t^{n}=27h$. The model faithfully reproduces the temperature profile. The wall model has a global error of  $\varepsilon_{\,\infty} \ = \ 0,36  \ \mathsf{^\circ C}$. 

The evolution of the simulated indoor air temperature is shown in figure \ref{fig:Comp_air}. As with the wall temperatures, the dynamics are well reproduced for all the models studied. The air temperature error is $0.20 \ \mathsf{^\circ C} $ .

\section{Model Validation with experimental data}
\label{sec:Practical application}

It is now proposed to study the behavior of the numerical model presented in a concrete case involving the simulation of a building in its urban environment, with realistic boundary conditions. The practical case chosen allows us to use all the models (Equations \ref{eq:heat_transfert_soil}, \ref{eq:heat_transfert_wall}, \ref{eq:inside_energy_balance}, \ref{eq:LW_in}, \ref{eq:LW_out}) developed in the micro-climatic simulation tool. Furthermore,  the model is assessed by comparing the experimental observations with the numerical predictions.

\subsection{Description}

\begin{figure}[ht]
    \centering
    \includegraphics[width=0.98\textwidth]{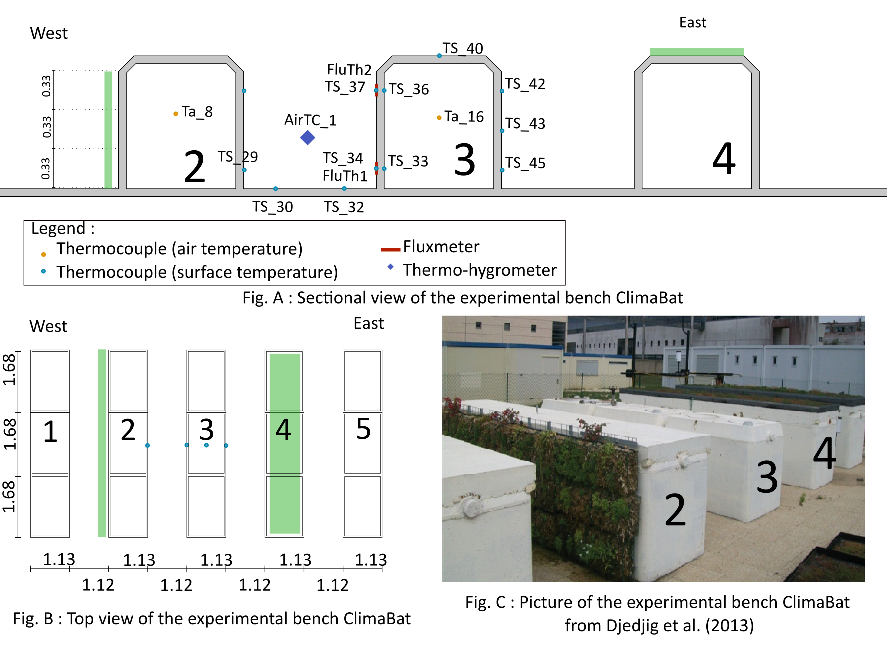}
    \caption{Location of thermocouples on the geometry studied.}
    \label{fig:Localisation_capteurs}
\end{figure}

For this practical case study, a geometric configuration representative of a real urban environment with experimental results has been chosen. The platform called ClimaBat \cite{djedjig2013experimental} was set up at La Rochelle university in France.  A first measurement campaign was conducted in 2009 to evaluate the performance of cool paintings \cite{doya2012experimental} . A second experimental campaign was conducted in 2012 with a new configuration: white painting on the building facade, two vegetated roofs and one vegetated facade.  Data from this second campaign are used for the model validation. The platform consists of a terrace ($10 \times 20$ m) of concrete slabs, on which empty concrete tanks have been arranged in several rows. They represent reduced-scale buildings (in the sense of small dimensions) composing several rows of buildings and streets as represented in Figure \ref{fig:Localisation_capteurs}). 

Access to reliable and complete experimental data at microclimatic scale to validate models under realistic conditions is a real challenge. For rigorous validation, most of the variables need to be mastered as the physical characteristics, the geometry, and the building usage. Most of the time, with real scale experiments, all those parameters are not known accurately as the studied sites are large (several buildings) and complex (in-use materials, occupants). A reduced scale experimental platform is an alternative. For model evaluation, this type of platform is an opportunity to control most parameters and reduce the complexity of the geometry considered. 

The geometry is the equivalent of a $1/10$-scale version of a 3-storey light-structure office building model (20 cm of hollow concrete blocks) surrounded by a pedestrian area. Each building row is $5.06 \, \mathsf{m}$ long (L), $1.24\, \mathsf{m}$ high (H) and $1.12\, \mathsf{m}$ wide (W). The width (W) of the street was set at $1.2$ m with an aspect ratio $H/W = 1.10$. The tanks consist of $e_w = 4$ to $5 \mathsf{cm}$ concrete walls \citep{doya2010analyse} with the following measured characteristics : $k_w \ = \ 2.35-2.47 \mathsf{W}. \mathsf{m}^{-1}. \mathsf{K}^{-1} $, $c_w = 2.27 - 2.3 \times 10^6 \ \mathsf{J}.\mathsf{m}^{-3}. \mathsf{K}^{-1}$. The floor is composed of $4 \, \mathsf{cm}$ thick concrete slabs with the following measured characteristics : $k_s \ = \ 1.90 - 2.19 \mathsf{W}. \mathsf{m}^{-1}. \mathsf{K}^{-1} $, $c_s = 1.946 \times 10^6 \ \mathsf{J}.\mathsf{m}^{-3}. \mathsf{K}^{-1}$. The concrete slabs laid on a $80 \, \mathsf{cm}$ of sand and gravel above the site natural soil. The thermal characteristics of the tanks and concrete slabs were measured during the thesis work of \cite{doya2010analyse}. The solar reflectivity (albedo) of the wall is $0.64$ and $0.36$ for the ground soil.  The characteristics of the ground soil below the concrete slabs were not measured. Standard values are therefore used for modeling purposes.

\subsection{Measurement set up}

The entire platform was instrumented to assess the performance of urban heat island mitigation solutions applied to buildings (plant walls, reflective paintings). Only the measurements used in the following sections are shown in Figure \ref{fig:Localisation_capteurs}. The experimental campaign is described in detail in \cite{doya2012experimental} and \cite{djedjig2013experimental}. Data from the experimental campaign of the first 24 days of August 2012 are considered for the investigations. 

Meteorological data measured on the roof ($16 \, \mathsf{m}$ height) of a university building are used as input data for the radiative model. The direct and diffuse components are calculated from the \cite{deMiguel2001} model, dividing the global irradiance measured by a pyranometer. 
The outside sensible heat flux (convection) is calculated from equation \ref{eq:h_c_out} using the temperature measured in the street by a thermo-hygrometer noted AirTC 1 on the Figure \ref{fig:Localisation_capteurs}. The measurement uncertainty of this sensor is estimated at $\sigma_{m}= \pm 0.4^\circ \, \mathsf{C}$. A 2D sonic anemometer is placed above the roof of the building row noted $3$ on the Figure \ref{fig:Localisation_capteurs}, at a height of $28 \, \mathsf{cm}$. Its measurement is subsequently used to calculate the convection coefficient with equation  \ref{eq:h_c_out}. 
\bigbreak
The tanks and slabs are equipped with thermocouples to measure surface temperatures at various points. The thermocouples used are type K with a sensor accuracy of $1.1 \ \mathsf{\degC}$ \cite{Th_info}. Finally, thermocouples inside the tanks measure the air temperature inside the central tank of each row. The measurement uncertainty of this sensor is also estimated at $\sigma_{m}= \pm 1.1^\circ \, \mathsf{C}$ \cite{Th_info}. 

\subsubsection{Geometry used: reduced or real size building ?}

As mentioned previously, the real case studied is a reduced-scale experimental demonstrator \cite{murphy_similitude_1950}. The tanks are standing for reduced-scale buildings made of hollow concrete blocks (20 cm).  Considering this point, the geometry modeled (real size or reduced size building) can be questioned. As this experiment data set has already been studied in several papers \cite{djedjig2013experimental, djedjig2015analysis}, this section justifies the choice of this case for validation of our model. 
% 1/  comment ces données ont été utilisées dnas l'EA
Work presented by \cite{djedjig2013experimental} studied the impact of vegetated roofs and facades on the urban microclimate based on the experimental results. Work presented by \cite{djedjig2012development} have used data set from another period (July 2012) to develop a green envelope model. This model is then integrated to a transient building simulation tool to investigate it effects on a multi-zones building in the work of \cite{djedjig2015analysis}. They proposed a brief comparison between the measured and simulated outside surface temperature for the reduced-scale geometry. The simulation is done using TRNSYS. Long-wave radiative heat fluxes exchanged between the building and their surrounding surfaces are not considered or simplified. The comparison (measured vs simulated data) is proposed for only two sensors with a RMSE of $2.75^\circ \ \mathsf{C}$ for the non-vegetated facades. Our work will complete this comparison between measured and calculated surface temperature, presenting the results for 14 sensors locations. The model is then used by \cite{djedjig2015analysis} on a three storey full scale building to transpose results to a realistic scale. 

As the objective of this paper is to have a rigorous comparison of modeled and measured temperature under realistic conditions, the question of geometry may arise: reduced or real-size building ? The parallel with life-size geometry has not been studied as a full similarities analysis has not been done. Under real urban conditions with real material, this point is very difficult to satisfy. According to \cite{doya2012experimental}, the studied reduced-scale experimental platform has the benefit to ensure the good reproduction of long-wave and solar radiation. This point has been verified by comparing the solar balance on real-size and reduced-size model using SOLENE simulation tool. The geometry of the experimental platform was designed to reduced wind velocities in the street and support the development of a thermal confinement in the street. Radiative heat flux are then the prevalent phenomenon studied.  However, the comparison between temperature measured at reduced scale and calculated with a model at real size proposed by \cite{doya2012experimental} is then questionable since the similarity laws with the real scale demonstrator are not respected. In other words, model validation with such experimental data can only be performed at the reduced-scale. 

\begin{figure}[ht]
    \centering
    \includegraphics[width=0.45\textwidth]{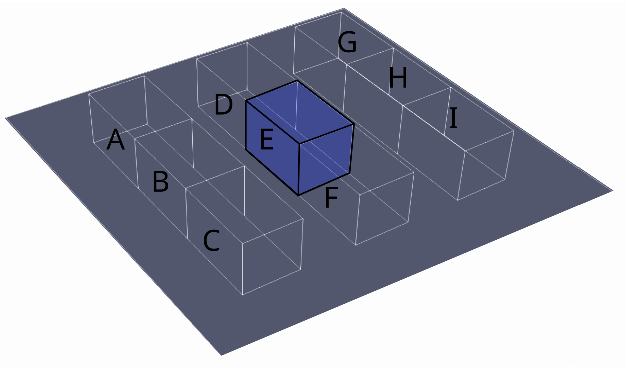}
    \caption{3D overview of the urban scene. }
    \label{fig:maquette}
\end{figure}

We decided then to model the platform with its original dimension and focus on the central part to stay away from vegetated areas. Figure \ref{fig:maquette} illustrates the geometry used. It is made up of the three central rows of buildings, as our study will focus on the behavior of the central building marked E (shown in light blue in the Figure). The green roof of buildings G, H and I and the green facade of buildings A, B, C is not modeled as it is not seen by building E surfaces directly.

\subsubsection{New model setup}

The thermal characteristics used are shown in Table \ref{tab:Caracteristiques_thermique}. A soil column of $e_s \ = \ 2 \mathsf{m}$ is considered in the model and a building wall thickness of $e_w \ = \ 4.5 \mathsf{cm}$.  Values of the solar reflectivity (albedo) measured are used for the walls ($0.64$) and for the ground soils ($0.36$). The emissivity is fixed at $0.9$ for each surface (it was not measured). The air volume (dimensions : $ 1.21 \ \times \ 1.68 \ \times \ 1.13 \ \mathsf{m}$ ) has a volumetric heat capacity $c_{air} \ = \ 1.2 \, 10^{\,3} \, \mathsf{J}. \mathsf{m}^{- 3}. \mathsf{K}^{-1}$ The initial temperature is set at $T_0 = 20 \ \mathsf{^\circ C}$ for all the material.
\begin{table}[ht]
    \centering
    \small
    \begin{tabular}{ccc}
    \hline\hline
        \textbf{Material}  &  \textbf{Heat conductivity} & \textbf{Volumetric heat capacity} \\
         &  $k \, [\mathsf{W}. \mathsf{m}^{-1}. \mathsf{K}^{-1} ] $ &  $c \, [ \mathsf{J}.\mathsf{m}^{-3}. \mathsf{K}^{-1}  ]$\\
    \hline
        Concrete of the tanks$^{1}$ & $2,37$ & $1.98 \times 10^6$ \\
        Concrete slabs$^{1}$ & $2,05$ & $1.95 \times 10^6$ \\
        Sand and gravel & $1.80$ & $1.40 \times 10^6$ \\
        Natural soil & $1.3$ & $1.44 \times 10^6$ \\
    \hline \hline
    \end{tabular}
    \caption{Thermal characteristics used for the simulation (Characteristics derived from measurements are indicated by this marker $^1$)}
    \label{tab:Caracteristiques_thermique}
\end{table}

On the outside of the urban scene, the flow balance includes convection and net radiation (SW and LW). The convection coefficient on the outside surface of the building enclosure is defined considering a linear variation with the measured wind speed $v_{air}$:
\begin{align}
\label{eq:h_c_out}
    h_{out} = 3.8 \times v_{air} + 5.0
\end{align}
The streets of the experimental setup were oriented North-South to protect them from the site prevailing winds, mainly from the West. To study the distribution of the velocity field near the buildings, \cite{doya2010analyse} carried out an aeraulic study of the site using the \texttt{ENVI-met} simulation tool. For a prevailing Westerly wind of $4 \  \mathsf{m}.\mathsf{s}^{-1}$, the wind speed in the street is low, and less than $0.3 \, \mathsf{m}. \mathsf{s}^{-1}$ at $0.4 \ \mathsf{m}$ height. This result confirms the use of Eq.~\eqref{eq:h_c_out} which is valid for wind speed under $5 \, \mathsf{m}/\mathsf{s}$ \citep{palyvos2008survey}. As the wind speed inside the street is very low, a fixed convection coefficient of $ h_{out} \ = \ 5 \ \mathsf{W}.\mathsf{m}^{-2}. \mathsf{K}^{-1} $ is used for the ground soil. 

On the interior faces of building E, in the same way, the flow balance includes convection and net radiation. Convection coefficients have been set according to the values prescribed by current thermal regulations \citep{RT-2012}: $ h_{in} \ = \ 0.7 \ \mathsf{W}.\mathsf{m}^{-2}$ for the floor, $ h_{in} \ = \ 2.5 \ \mathsf{W}.\mathsf{m}^{-2}$ for the internal walls surface and $ h_{in} \ = \ 5 \ \mathsf{W}.\mathsf{m}^{-2}$ for the roof internal surface. As the tanks are opaque, only the long-wave radiative flux is considered for the net radiation balance. The emissivity of the tank is fixed at $0.9$. 
\bigbreak
Building E has an adjoining wall with each of buildings D and F. The thickness of the adjoining walls is considered to be double (the tanks are placed next to each other). The contact resistance between the two tanks is neglected. To simplify the calculation, the air temperature in the two adjoining buildings is assumed to be equal to the measured air temperature in the building under study. 
\bigbreak
For the other buildings (A, B, C, D, F, G, H, I), the inside air temperature is fixed to the measured temperature. In the ground soil, the boundary condition at a depth of $2$ m in both the exterior and interior ground soil is considered constant, and is set at the mean air temperature of $22.7 \ \mathsf{\degC}$. This parameter was not measured.
\bigbreak
The simulation is run from the 01/08/2012 00:00 to the 24/08/2012 23:45. The simulation time step has been set to $900 \, \mathsf{s}$ and the spatial mesh uses $\texttt{N}_x = 101$ nodes. 

\subsubsection{SOLENE-Microclimat setup}

\texttt{SOLENE-microclimat} is used with the thermo-radiative function originally developed \cite{musy2015use}. For both models, the same setup is used: material properties, outside convection coefficient calculation, simulation time-step and period, short-wave calculation, geometry and mesh. As the tool does not allow having the exact same configuration for some points, the following differences should be noted: (1) in the ground soil the boundary condition is set at a depth of $2$ m with a temperature of $17.7 \ \mathsf{\degC}$, (2) in the building the temperature is constant and set as the mean temperature measured in the building E \textit{i.e.} $25.7 \ \mathsf{\degC}$, (3) inside convection coefficients are set constant. 

\subsubsection{Local sensitivity analysis}
For both models, the values used are either taken from the literature in the absence of exact data provided by the original work of the experimental campaign. 
No attempt has been made to calibrate the model predictions with the experimental data. However, to evaluate the influence of uncertain parameters on the model predictions, a local sensitivity analysis is performed. Two main parameters are chosen due to their large uncertainty and influences on the predictions: the concrete wall length $e_w$ and the urban environment albedo noted $\alpha$ . For this, a \textsc{Taylor} expansion of the model solution is performed \cite{jumabekova_efficient_2021}:
\begin{align*}
    T(x\,,\,t\,,\,\alpha\,,\,e_w\,) \,=\,
    T(x\,,\,t\,,\,\alpha_{\,0}\,,\,e_{\,w, \, 0}\,) \,+\, &
    \bigl(\, \alpha \,-\, \alpha_{\,0}\, \bigr) \, \frac{\partial T}{\partial \alpha} \,+\, 
    \bigl(\, e_w \,-\, e_{\, w , \,0}\, \bigr) \, \frac{\partial T}{\partial e_w}\\[4pt]
    & \,+\, \mathcal{O}\,\Bigl(\,\bigl(\, \alpha \,-\, \alpha_{\,0}\, \bigr)^{\,2}\,,\,\bigl(\, e_w \,-\, e_{\, w, \,0}\, \bigr)^{\,2}\,\Bigr) \,,
\end{align*}
where $\alpha_{\,0}$ and $e_{\, w, \,0}$ are the reference values for albedo and concrete wall length. The partial derivative, denoted as sensitivity functions, $\frac{\partial T}{\partial \alpha} $ and $\frac{\partial T}{\partial e_w} $ are computed using a central finite-differences approximation of order 1. The upper and lower values of albedo and length considered for the expansion are given in Table~\ref{tab:local_SA_values}.

\begin{table}[ht]
    \centering
        \small
    \begin{tabular}{c ccc}
    \hline \hline
     & reference & upper & lower \\\hline 
    $e_w \ [\,\mathsf{cm}\,]$ &  $4.5$  & $5$ & $4$\\ 
    $\alpha \ [\,-\,]$ & $0.64$ & $0.6$ & $0.69$ \\ \hline \hline
    \end{tabular}
    \caption{Numerical values considered for the local sensitivity analysis.}
    \label{tab:local_SA_values}
\end{table}

\subsection{Results}

This case study is used to validate the model predictions with experimental data in realistic conditions. The models are evaluated on their ability to reproduce the dynamics of realistic solicitations. Results of the model are compared to measurement and to \texttt{SOLENE-microclimat} simulation tool as a reference tool in the field.

\subsubsection{Heat fluxes balance }

\begin{figure}[htp]
    \begin{minipage}[c]{.46\textwidth}
  \centering
  \subfigure[Lower part of the wall  $FluTh1$]{
  \label{fig:FluTh1}
  \includegraphics[width=1\textwidth]{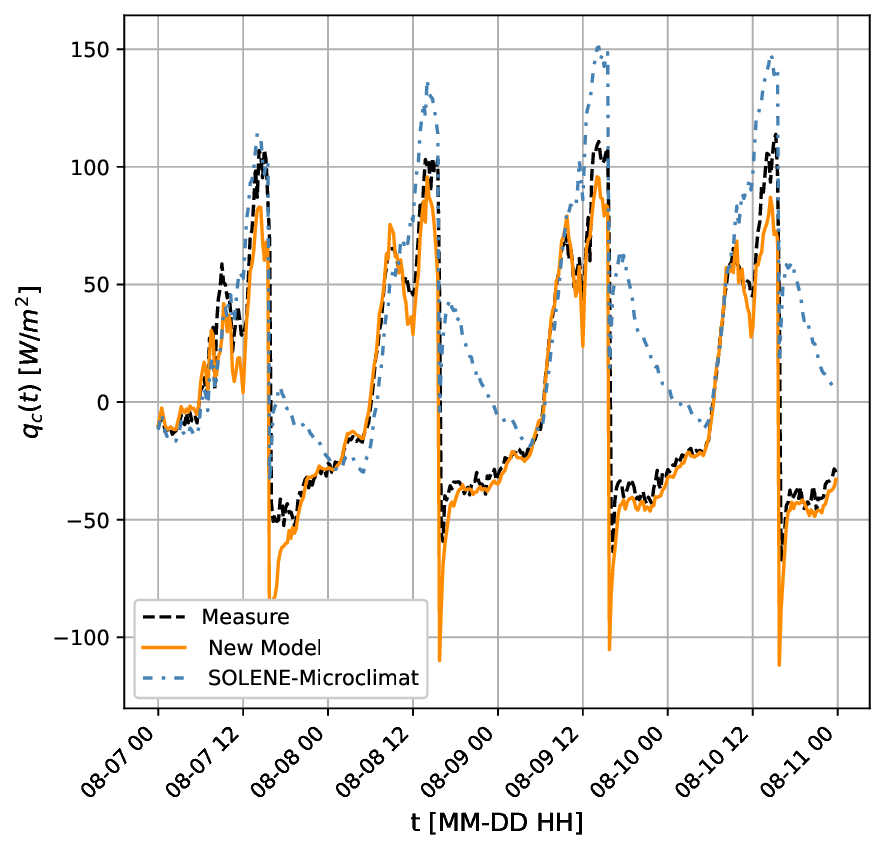}
  }
    \end{minipage}
        \hfill%
    \begin{minipage}[c]{.46\textwidth}
  \subfigure[Upper part of the wall  $FluTh2$ ]{
  \label{fig:FluTh2}
  \includegraphics[width=1\textwidth]{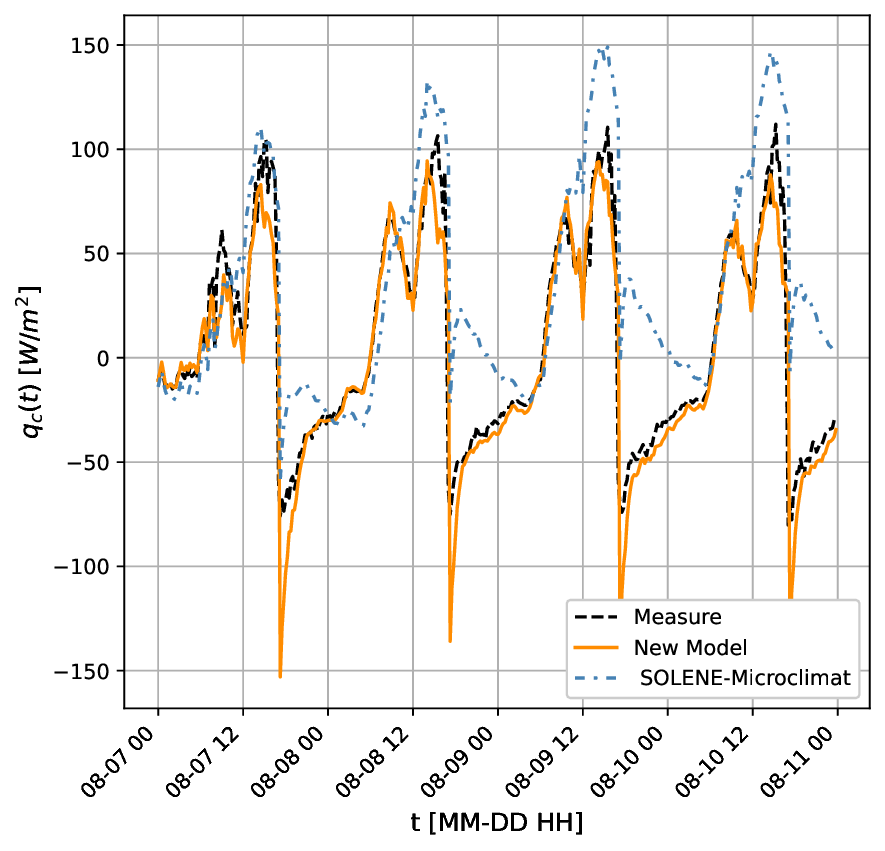}
  }
    \end{minipage}
  \caption{Comparison of West wall conductive flux density measured and calculated by the models.}
  \label{fig:flux_conduit}
\end{figure}

As the main objective of this work is to improve inside and outside long-wave radiative balance, the heat budget is first compared. Figure \ref{fig:flux_conduit} shows the comparison between the calculated and measured conduction flux density in the West wall (upper and lower part). The new model correctly calculates the heat balance at the surface. The new conductive heat flux in the wall follows the measured one with good accuracy. The new model shows improvement from \texttt{SOLENE-microclimat} original model at noon and during the night with an error of $21,19 \ W/m^{2}$ against $34,48 \ W/m^{2}$ for the lower part of the wall and $22,01 \ W/m^{2}$ against $37,47 \ W/m^{2}$ for the upper part of the wall. 

\begin{figure}[htp]
    \begin{minipage}[c]{.46\textwidth}
  \centering
  \subfigure[Long-wave radiative fluxes]{
  \label{fig:flux_GLO}
  \includegraphics[width=1\textwidth]{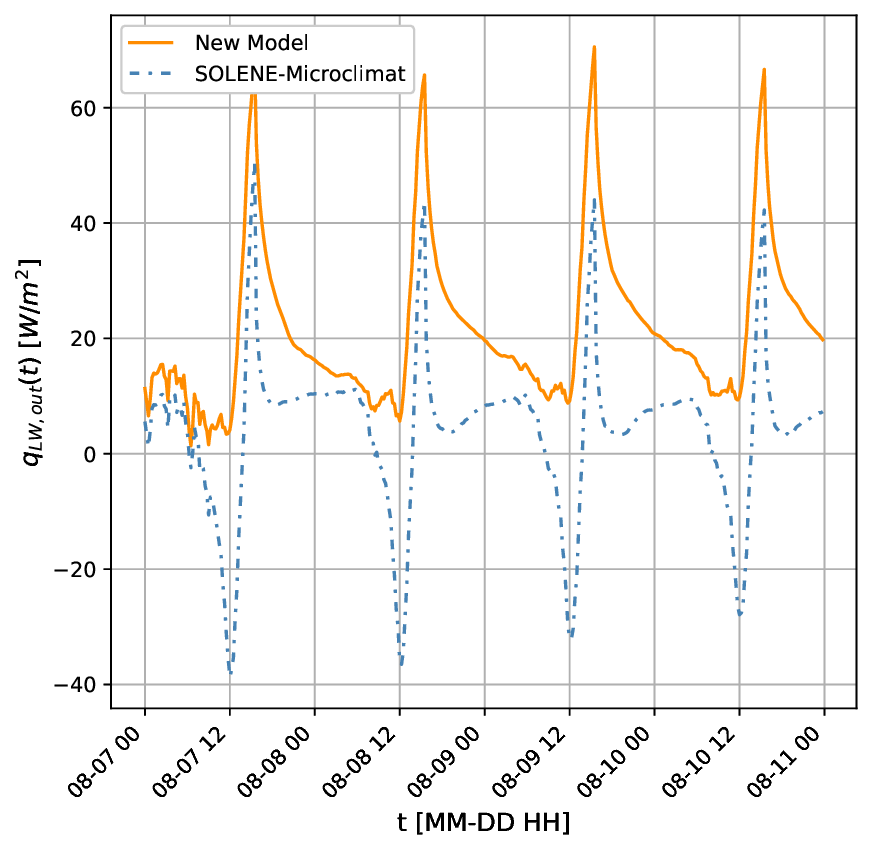}
  }
    \end{minipage}
        \hfill%
    \begin{minipage}[c]{.46\textwidth}
  \subfigure[Short-wave radiative flux and convective heat flux]{
  \label{fig:flux_CLO_h}
  \includegraphics[width=\textwidth]{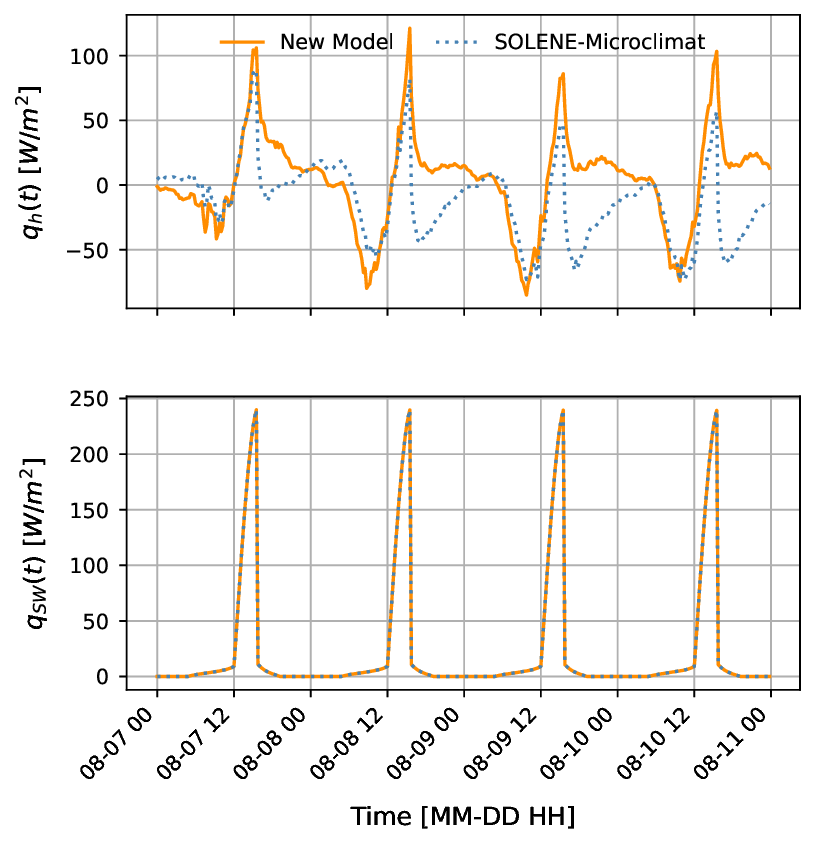}
  }
    \end{minipage}
  \caption{Comparison of heat flux density calculated by both models.}
  \label{fig:heat_flux_budget}
\end{figure}

This difference is partly due to long-wave radiative heat fluxes calculation. Figure \ref{fig:flux_GLO} shows the difference between the two models for the lower part of the wall. The same tendency can be observed for the upper part of the wall. This difference between the two curves can be linked to the new form factor calculation, the radiosity model, and the conduction model in the wall (as it is linked to the surface temperature).  \texttt{SOLENE-microclimat} building wall model assumes a RC lumped capacitance model, which is known to be less accurate to represent the physical phenomena \cite{berger_comparison_2020,berger_evaluation_2019}. 

The form factor calculation method implemented in SOLENE simplifies the calculation by \textbackslash{}emph\{(i)\} integrating over the contours of the two polygons that enclose the surfaces (rather than calculating the double integral over the two polygons concerned) and \textbackslash{}emph\{(ii)\} the emitting polygon is considered as an infinitesimal surface, \textbackslash{}emph\{(i.e.)\} a point source \textbackslash{}cite\{miguet2000parametres\} . The form factor is therefore calculated from a point to a polygon and not from polygon to polygon (as it is in the new method implemented here). This simplification remains valid as long as the distance to the receiving surface is greater than around 5 times the maximum dimension of the projected emitting surface. Obviously, this condition is not met for two adjacent, non-coplanar surface elements. It is here problematic in corners of a canyon street geometry. 

The new model underestimates the surface heat flux balance around $16:00$ each day. Since the short-wave radiative heat flux is the same for both models, as shown in Figure \ref{fig:flux_CLO_h}, and since the convection coefficients are modeled similarly, the difference between the surface temperature and the air temperature is further investigated. At this moment of the day, the surface temperature of the model decreases faster than the measured one and reaches a difference of about $10 \ \mathsf{\degC}$ with the air temperature. As noted in the top of Figure \ref{fig:flux_CLO_h}, the convective heat flux value is then overestimated. 

\subsubsection{Building zone E modeling}

The validation of the building model is now assessed through the comparison of the surface and air temperature variations. The errors with respect to experimental data for surface and air temperatures for both models are given in Table \ref{tab:Err_exp_batiment_LR}. The errors obtained are under the one from the literature. \cite{djedjig2015analysis} obtained a RMSE of $2.75^\circ \ \mathsf{C} $ for a sensor located on the west wall where the new model calculated error is $1.87^\circ \ \mathsf{C}$.

\begin{table}[ht]
    \centering
    \small
    \begin{tabular}{ccccccccc}
        \hline\hline
        \textbf{Sensors}  &  $TS \, 36$ & $TS \, 37$ & $TS \, 33$ & $TS \, 34$ & $TS \, 45$ & $TS \, 42$ & $Ta \, 16$ & $TS \, 40$  \\ \hline
        \textbf{Location}  &  in, up & out, up& in, down& out, down& out, down & out, up& in & out  \\ 
          &  W. wall & W. wall & W. wall & W. wall & E. wall & E. wall & Air & Roof  \\ %\hline        
        \textbf{New model } & $1,72$ & $1,86$ & $2.13$ & $1.87$ & $2.16$ & $2.76$ & $1,72$ & $2,31$ \\ %\hline
        \textbf{\texttt{SOLENE-micro.} }  & - & $3,06$ & - & $2,81$ & $3,36$ & $4,38$ & - & $3,93$\\ \hline \hline
        
    \end{tabular}
    \caption{Errors $\varepsilon_2$  with respect to experimental data for the all simulation period (out from the 6 first initialization days)}
    \label{tab:Err_exp_batiment_LR}
\end{table}

\begin{figure}[htp]
    \begin{minipage}[c]{.46\textwidth}
  \centering
  \subfigure[Inside surface temperature $TS \, 36$]{
  \label{fig:TS_36}
  \includegraphics[width=1\textwidth]{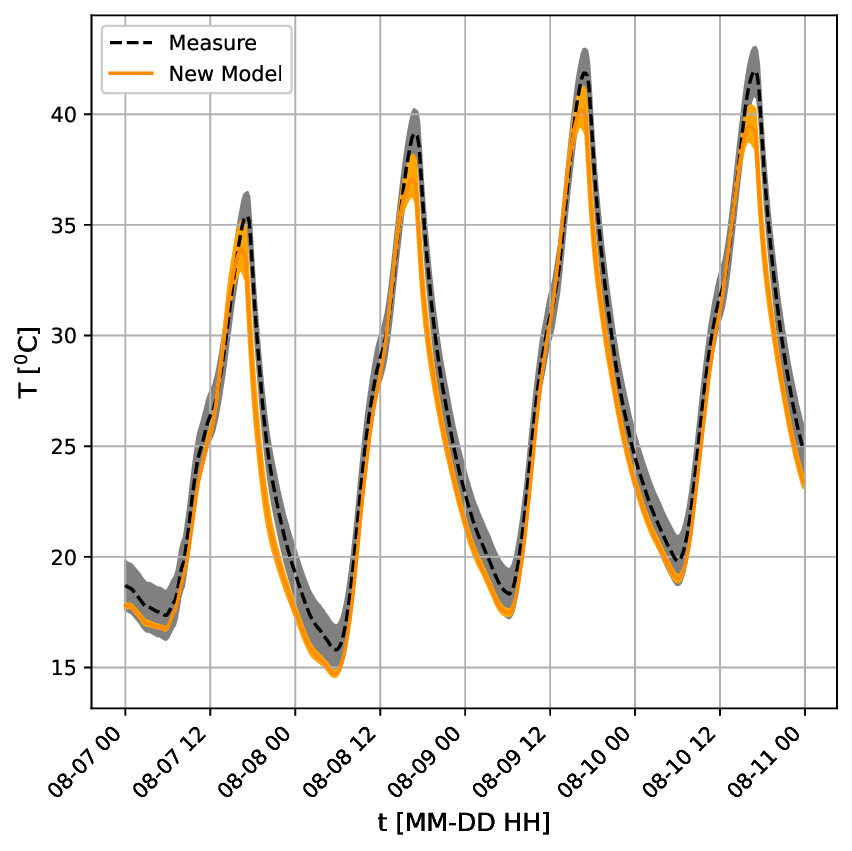}
  }
    \end{minipage}
        \hfill%
    \begin{minipage}[c]{.46\textwidth}
  \subfigure[Outside surface temperature $TS \, 37$ ]{
  \label{fig:TS_37}
  \includegraphics[width=1\textwidth]{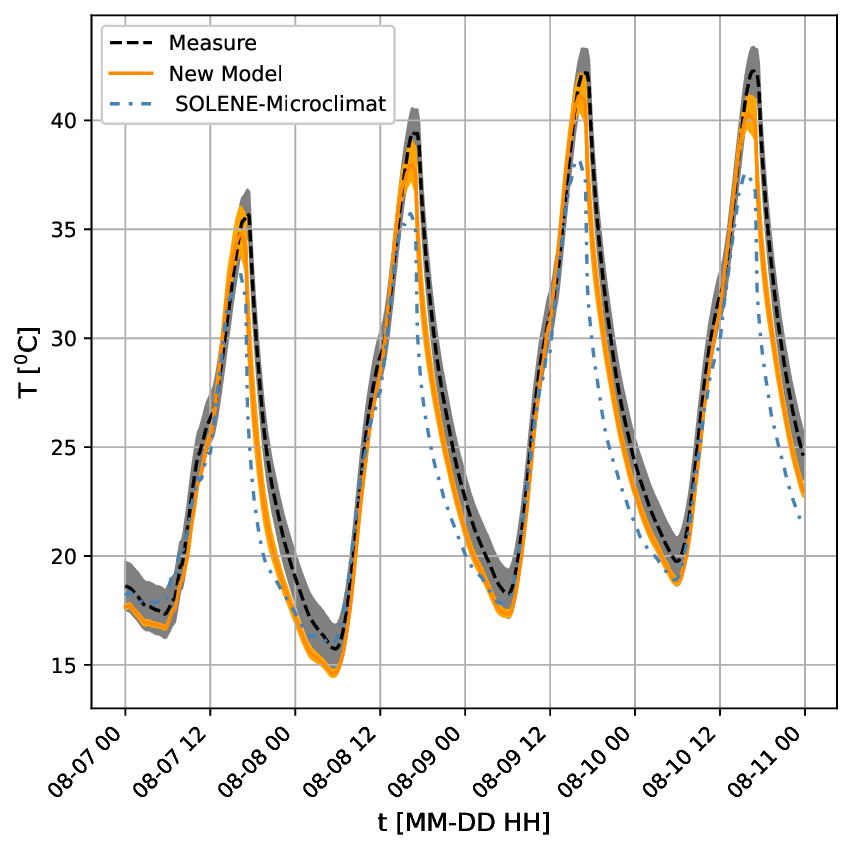}
  }
    \end{minipage}
        \vfill%
    \begin{minipage}[c]{.46\textwidth}
  \centering
  \subfigure[Inside surface temperature calculated by the new model in Kelvin, the 08/08 at 13:15. Surface of interest are highlighted. ]{
  \label{fig:TS_int}
  \includegraphics[width=1\textwidth]{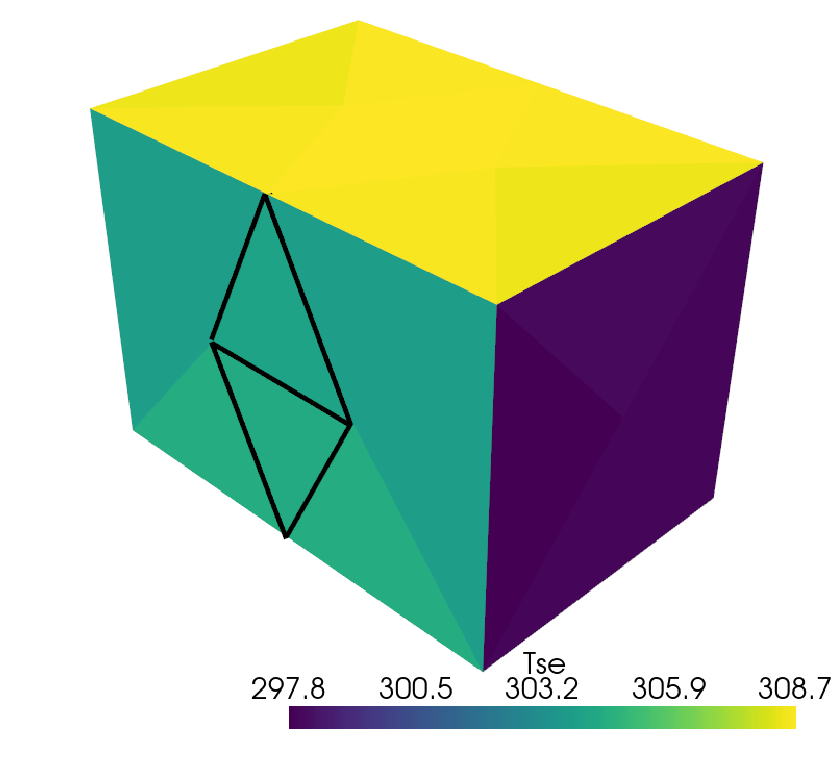}
  }
    \end{minipage}
        \hfill%
    \begin{minipage}[c]{.46\textwidth}
  \subfigure[Outside surface temperature $TS \, 34$ ]{
  \label{fig:TS_34}
  \includegraphics[width=1\textwidth]{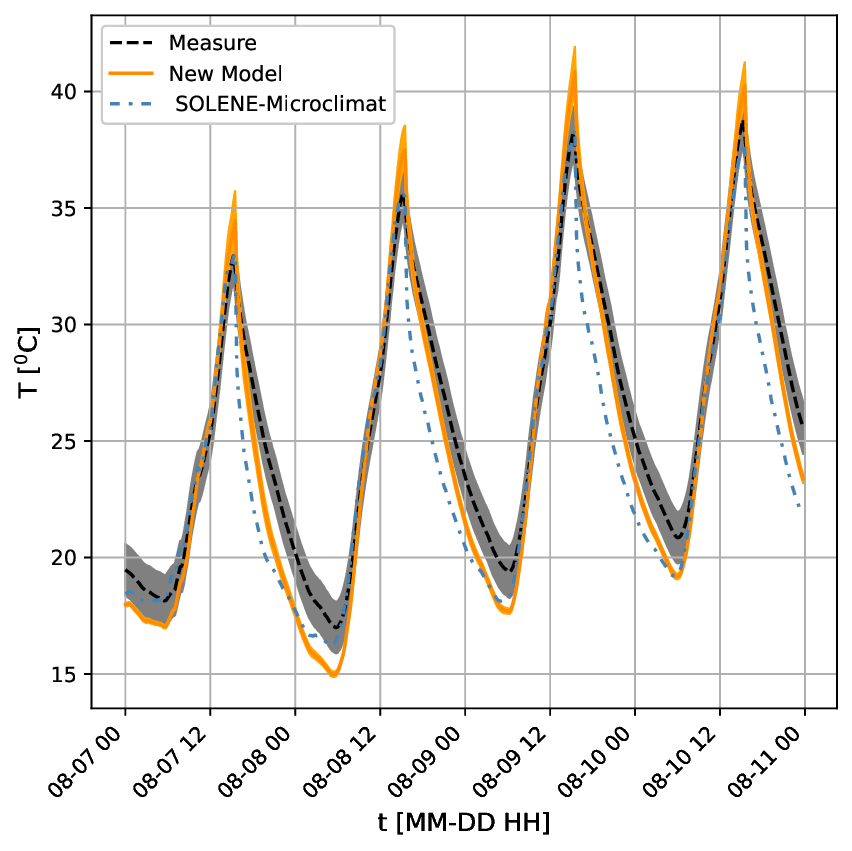}
  }
    \end{minipage}
  \caption{Comparison of West wall surface temperatures measured and calculated by the models.}
  \label{fig:comp_temp_surf_ouest}
\end{figure}

Figure \ref{fig:comp_temp_surf_ouest} shows the evolution of surface temperatures on the West wall.  Figures \ref{fig:TS_37} and \ref{fig:TS_34} present the exterior surface temperatures. The new model faithfully reproduces the dynamics. It reduces the error by $1.18 \ \mathsf{\degC}$ over \texttt{SOLENE-microclimat} initial error for the upper part of the wall and by $0.89 \ \mathsf{\degC}$ for the lower part of the wall. 

Figures \ref{fig:TS_36} and \ref{fig:TS_37} show the building interior surface temperatures. Building inside surfaces are not discretized in \texttt{SOLENE-microclimat}. A surface temperature is calculated per face of a zone.  It explains why Figure \ref{fig:TS_36} only presents results from the new model. In the new model, the inside heat balance is improved.  Surface temperature results are shown on the Figure \ref{fig:TS_int} for a time step. It can be seen in the Figure \ref{fig:TS_36}, the new model faithfully reproduces the dynamics and underestimates the inside surface temperature. 

\begin{figure}[htp]
    \begin{minipage}[c]{.46\textwidth}
  \centering
  \subfigure[Outside surface temperature $TS 42$]{
  \label{fig:TS_42}
  \includegraphics[width=1\textwidth]{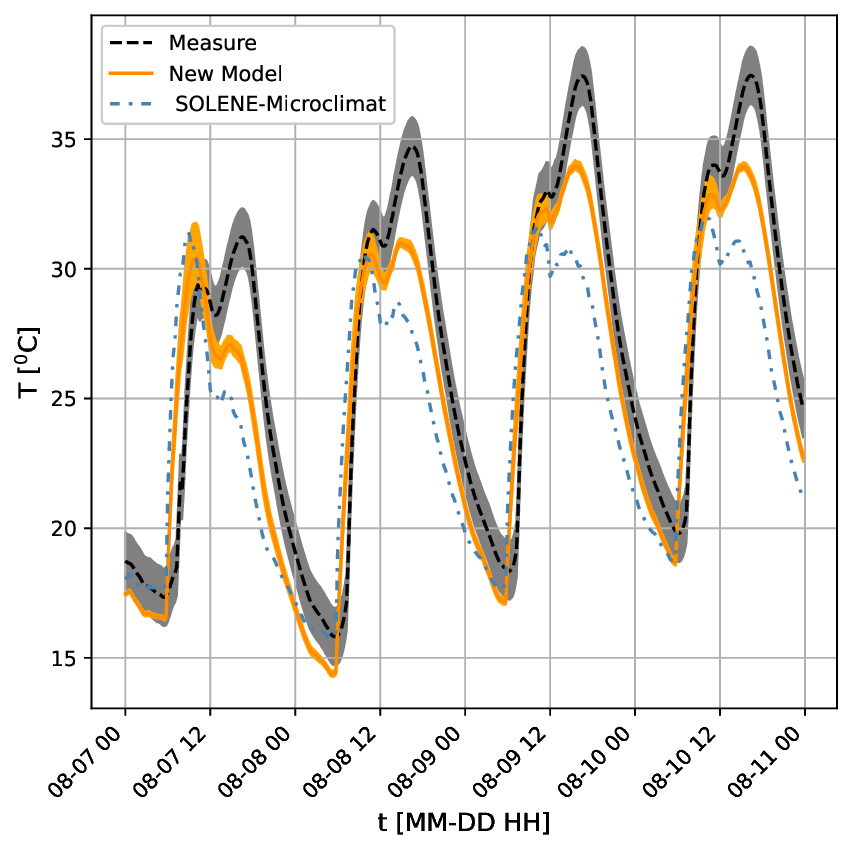}
  }
    \end{minipage}
        \hfill%
    \begin{minipage}[c]{.46\textwidth}
  \subfigure[Outside surface temperature $TS 45$ ]{
  \label{fig:TS_45}
  \includegraphics[width=1\textwidth]{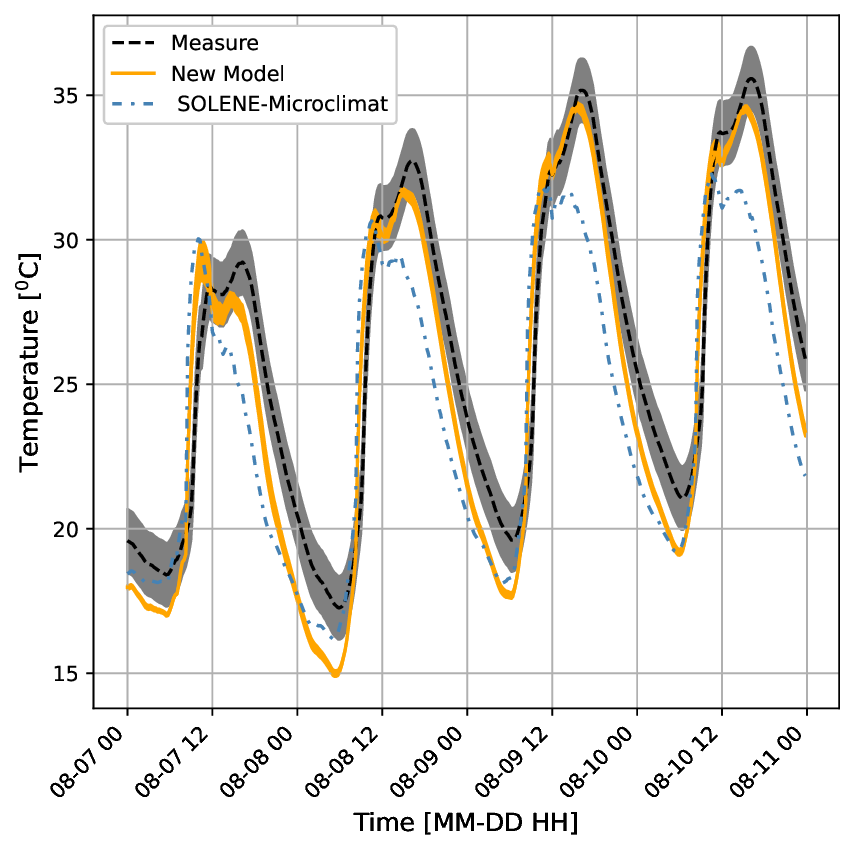}
  }
    \end{minipage}
  \caption{Comparison of East wall surface temperatures measured and calculated by the models.}
  \label{fig:comp_temp_surf_est}
\end{figure}

Figure \ref{fig:comp_temp_surf_est} shows the evolution of the external surface temperature of the East wall. The calculated surface temperatures are more distant from the surface temperature measured during the day. Both models underestimate the surface temperature. For the new model, an error in the parameters affecting the short-wave radiation balance or the sensible heat flux can be the cause of this discrepancy.

\begin{figure}[h!]
    \centering
    % Première figure
    \begin{minipage}{0.45\textwidth}
    \centering
    \includegraphics[width=\textwidth]{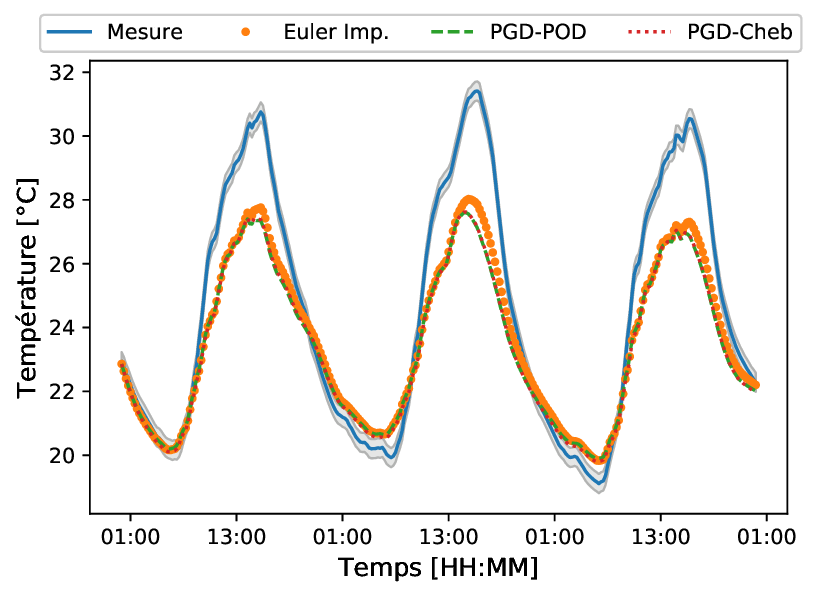}
    \caption{Comparison of the indoor air temperature in Building E measured and calculated by the model}
    \label{fig:comp_temp_air}
    \end{minipage} \hfill
    % Deuxième figure
    \begin{minipage}{0.45\textwidth}
    \centering
    \includegraphics[width=\textwidth]{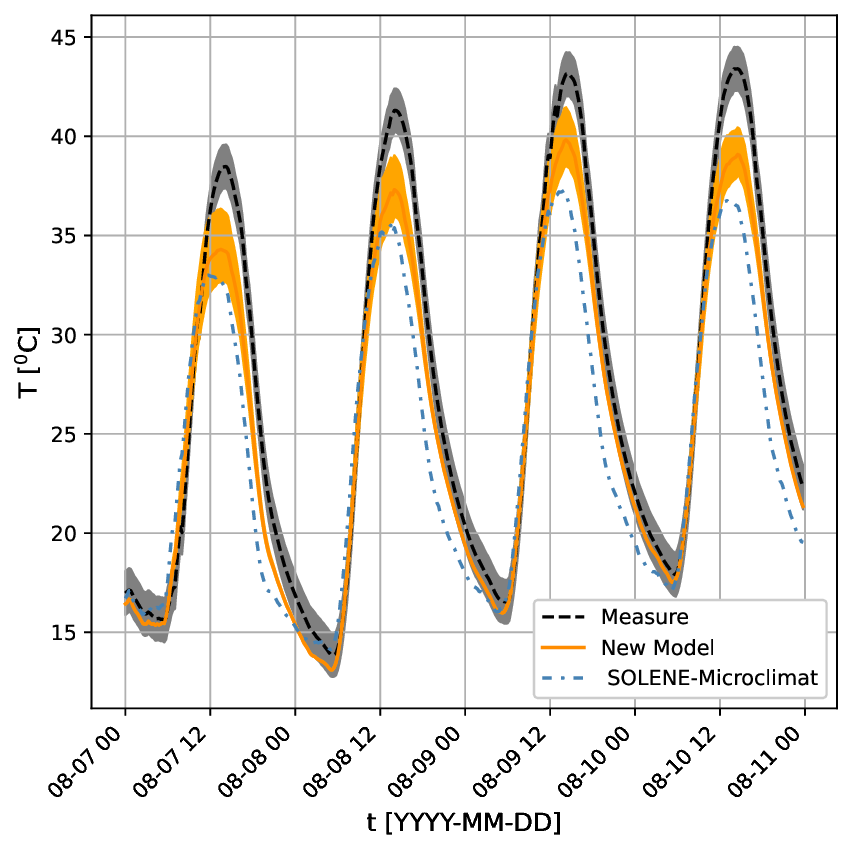}
    \caption{Exterior roof surface temperature comparison.}
    \label{fig:comp_temp_toit}
    \end{minipage}
\end{figure}

Figure \ref{fig:comp_temp_air} shows the evolution of the air temperature inside the building for the total period of simulation (out from the 6 first initialization days). The model underestimates the indoor air temperature which can be allocated to a consequence of the minor errors on the surface wall computations. As the temperatures are mainly underestimated for the various surfaces, the air temperature is also underestimated. The calculated error is $1.72  \ \mathsf{\degC}$.

Results of the temperatures calculated on the roof are presented in Figure \ref{fig:comp_temp_toit}. For both models, the dynamics are well reproduced, but surface temperatures are underestimated during the day. The error increases for \texttt{SOLENE-microclimat} mainly during the day. For both models, a clear weather day is more difficult to model. As previously said, the short-wave radiation balance can be the cause of this discrepancy, since the error is mainly observed during the day with high short-wave radiative fluxes. A local sensitivity analysis of the albedo value is performed and indicated by the orange shadow in Figure \ref{fig:comp_temp_toit}. Increasing the albedo value reduces the discrepancy between the model predictions and the experimental data. An error of $3.93 \ \mathsf{\degC}$ is calculated for \texttt{SOLENE-microclimat} and $2.31  \ \mathsf{\degC}$ for the new model.

\subsubsection{Outdoor urban ground soil and building modeling}

\begin{figure}[htp]
    \begin{minipage}[c]{.46\textwidth}
  \centering
  \subfigure[Outside ground soil surface temperature $TS 30$]{
  \label{fig:TS_30}
  \includegraphics[width=1\textwidth]{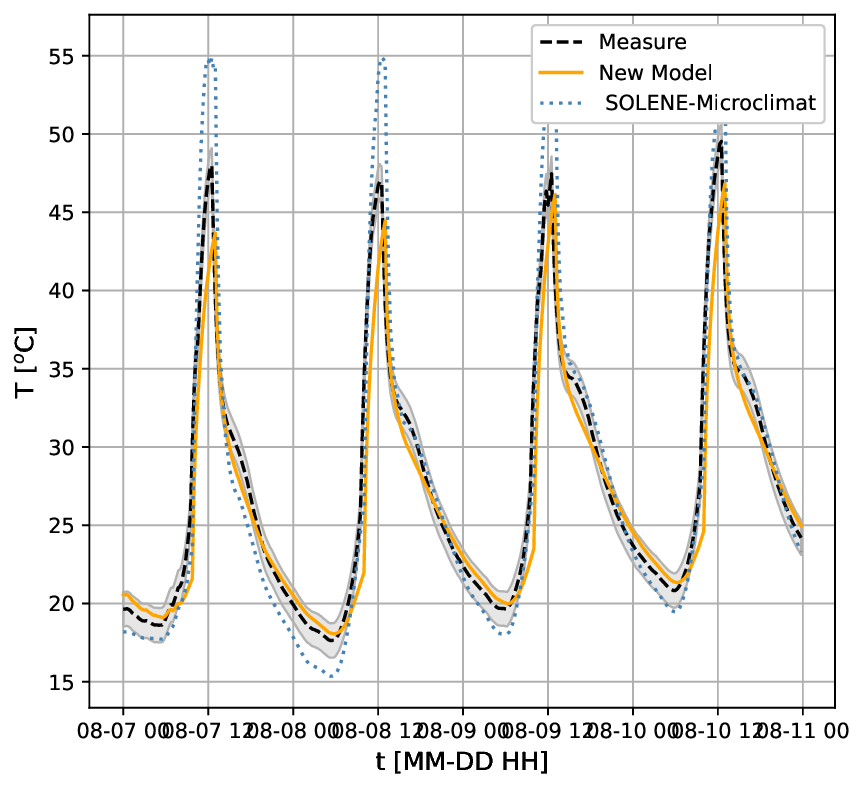}
  }
    \end{minipage}
        \hfill%
    \begin{minipage}[c]{.46\textwidth}
  \subfigure[Outside surface temperature calculated by the new model, the 08/08 at 13:15. Surface of interest are highlighted.]{
  \label{fig:TS_out}
  \includegraphics[width=1\textwidth]{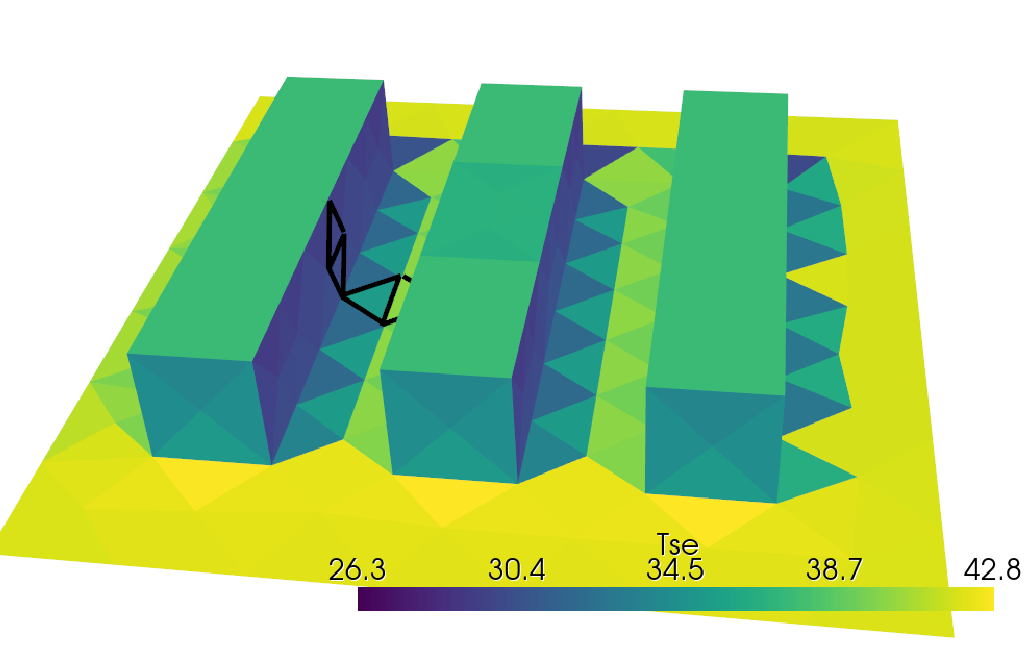}
  }
    \end{minipage}
    \begin{minipage}[c]{.46\textwidth}
  \centering
  \subfigure[ Outside building surface temperature TS 26]{
  \label{fig:TS_26}
  \includegraphics[width=1\textwidth]{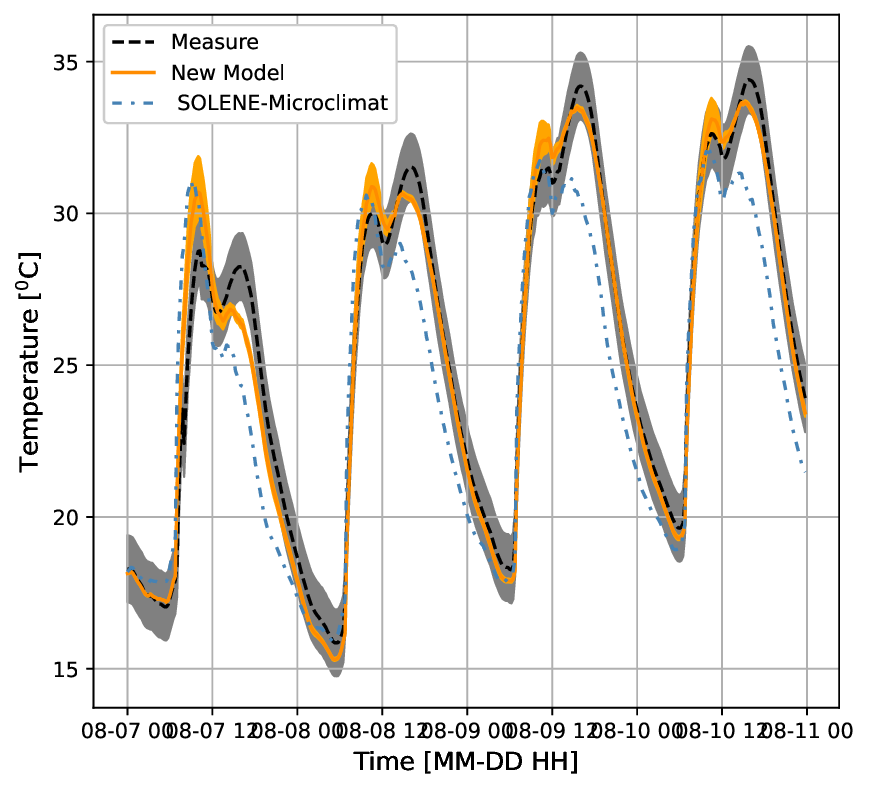}
  }
    \end{minipage}
        \hfill%
    \begin{minipage}[c]{.46\textwidth}
  \subfigure[ Outside building surface temperature TS 29 ]{
  \label{fig:TS_29}
  \includegraphics[width=1\textwidth]{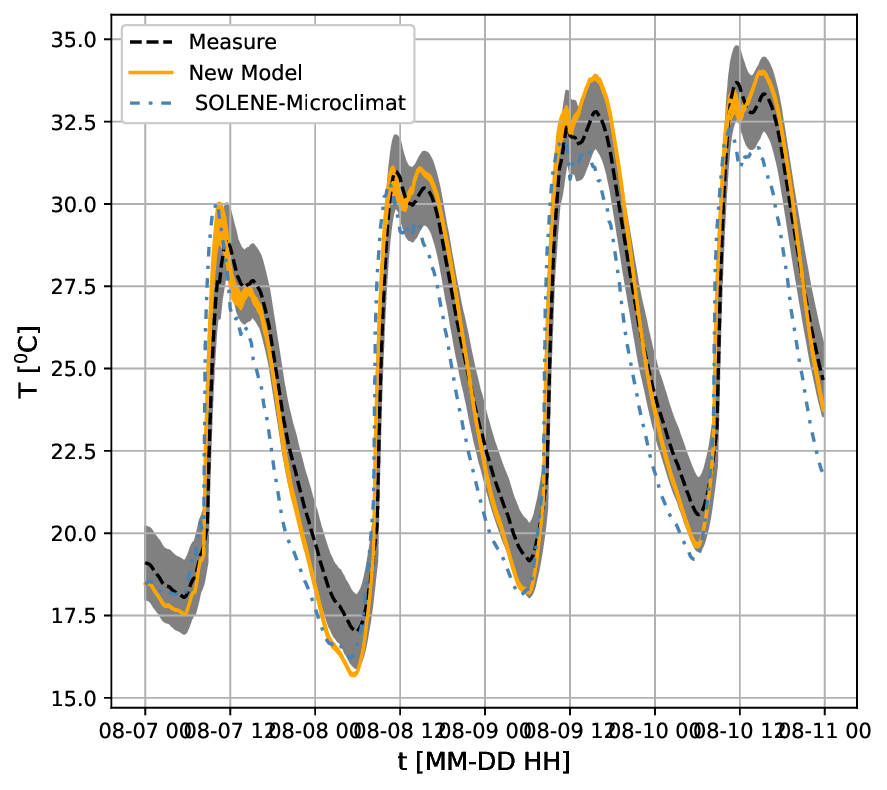}
  }
    \end{minipage}
  \caption{Comparison of outdoor surface temperatures measured and calculated by the models.}
  \label{fig:comp_temp_surf_sol}
\end{figure}

For the all urban scene, surface temperature and energy balance are calculated. Figure \ref{fig:TS_out} presents the outside surface temperature calculated by the new model, the 08/08 at 13:15. The surfaces of interest are highlighted in black in the Figure \ref{fig:TS_out}. Out from the central building zone, the outside surfaces are either impervious ground soils or building components (walls or roofs). 

Modeling outside urban ground soil is not easy, as very little information is known on its composition on site. Only the first four centimeters have been characterized. Even if the same material properties have been considered in both models, as can be seen in Figure \ref{fig:TS_30}, the numerical models have two different behaviors. \texttt{SOLENE-microclimat} overestimates the surface temperature. This is a known default of the model, which has been corrected in recent development \cite{azam2017new}. The proposed model here is more detailed and underestimates daytime outdoor surface temperatures during the day. This error is probably mainly due to unknown ground soil characteristics. 

For the building surface, Figures \ref{fig:TS_26} and \ref{fig:TS_29} show the results. Even if the physics is less detailed for those surfaces (simplification of the long-wave calculation inside and imposed air temperature) the new model reproduces well the dynamics with errors on the same magnitude as for the previous results $1.26  \ \mathsf{\degC}$ to $1.57 \ \mathsf{\degC}$. There is a shift in the calculated surface temperature by \texttt{SOLENE-microclimat} models, which leads to a larger error  $2.45 \ \mathsf{\degC}$ to $2.79 \ \mathsf{\degC}$.

\begin{table}[htp]
    \centering
    \small
    \begin{tabular}{ccccc}
    \hline \hline
    \textbf{Sensors} & TS 30 & TS 32 & TS 26 & TS 29 \\ \hline
    \textbf{Location} & G. Soil & G. Soil & Wall, out, up & Wall, out, down \\ %\hline
        \textbf{New model}  & $3,02$ & $3,51$ & $1,57$ & $1,26$ \\
 %\hline
    \textbf{\texttt{SOLENE-microclimat} }  & $4,62$ & $5,58$ & $2,79$ & $2,45$ \\ \hline \hline

    \end{tabular}
    \caption{Errors $\varepsilon_2$ with respect to the experimental data. }
    \label{tab:erreur_sol_exp}
\end{table}

\section{Conclusion}
This work presents a new microclimatic model at the district scale with improvement of long-wave radiative heat balance outside and inside of the urban scene. To achieve this point (1) two geometries that coincides are involved : one for the urban scene and one for the studied zone (2) a new form factor calculation is implemented (inside and outside) with the radiosity method.

The new models have been validated with a validation case studies with a reference solution. The new microclimate model is then applied to an urban scene with realistic solicitation. For that purpose, data from an experimental demonstrator presented in the litterature was used. The assessment has been conducted on 14 sensors (temperature and heat fluxes) over 24 days period on various surfaces (walls, roofs, ground soils). Those comparisons with experimental data have demonstrated the ability of the model to reproduce the dynamics of realistic solicitation with a RMSE between $1.26$ and $2.76  \ \mathsf{\degC}$ for the building surface temperature and a maximum RMSE of $3.51  \ \mathsf{\degC}$ for the ground soil. 

Results of the new model have also been compared to ones from state-of-the-art simulations tool. It shows improvement on the surface energy balance as the long-wave radiative balance is better modeled and on the surface temperature calculation as a consequence. The new model presents a better reliability with the experimental observations of $0.9 \ \mathsf{\degC}$ to $2.1 \ \mathsf{\degC}$ compared to the reference simulation tool for the specific case study.

Improving the physics representation on those types of models opens opportunities to study the effect of urban mitigation scenarii on the inside building conditions: air temperature modeling during a heat wave, air conditioning energy consumption evaluation. 

\newpage 
\appendix

\section{Dimensionless Equations}
\label{Appendix1}
\textbf{Heat transfer in the building wall}\\ 
\bigbreak
The problem can be rewritten in the following dimensionless form: 
\begin{equation}
c_w(x)^{\,\star}  \frac{\partial T^{\,\star}}{\partial t ^{\,\star} } \ = \ Fo (x) \frac{\partial}{\partial  x^{\,\star}} \, \biggl(\, k_w (x)^{\,\star} \, \frac{\partial T^{\,\star} }{\partial x^{\,\star}} \, \biggr) \,,
 \label{Eq:EqChaleurAdim}
\end{equation}
and the boundary conditions as :
\begin{subequations}
\label{eq:bc_dimless}
\begin{align}
\ k^{\,\star}  \frac{\partial T^{\,\star} }{\partial x^{\,\star}}   &  = \  Bi_{out} \, \Bigl(\, T^{\,\star} \ - \ T_{\,out}^{\,\star} \, \Bigr) \ - \ q_{net, \, out}^{\,\star} \,, && x^{\,\star}  \ = \  0 \,,  \\[4pt]
\ k^{\,\star} \frac{\partial T^{\,\star} }{\partial x^{\,\star}}   &  = \ q_{net, \, in}^{\,\star} \ - \  Bi_{in} \, \Bigl(\, T^{\,\star} \ - \ T_{\,in}^{\,\star} \, \Bigr)\,, && x^{\,\star} \ = \ 1 \,, \\[4pt]
T^{\,\star} & \ = \  0,  && t^{\,\star} \ = \ 0\, .
\end{align}
\label{Eq:BC_Adim_mur}
\end{subequations}
\bigbreak
\textbf{Heat transfer in the ground soil}\\ 
\bigbreak
The problem can be rewritten in the same dimensionless form as in \eqref{Eq:EqChaleurAdim} with the following boundary conditions :
\begin{subequations}
\label{eq:bc_dimless2}
\begin{align}
\ k^{\,\star} \frac{\partial T^{\,\star} (x,t) }{\partial x^{\,\star}}   &  = \  Bi_{out} \, \Bigl(\, T^{\,\star} \ - \ T_{\,out}^{\,\star} \, \Bigr) \ - \ q_{net, \, out}^{\,\star} \,, && x^{\,\star}  \ = \  0 \,,  \\[4pt]
\   T^{\,\star} (x,t)  & \ = \ T^{\,\star}_{\infty},  && x^{\,\star} \ = \  1 \,, \\[4pt]
T^{\,\star} & \ = \  0,  && t^{\,\star} \ = \ 0\, .
\end{align}
\label{Eq:BC_Adim_sol}
\end{subequations}

The two previous systems of equations use the following dimensionless numbers :
\begin{equation*}
T^{\,\star} : \ = \ \frac{T \ - \ T_{\,0}}{T_{\,0}} ; \; \; \;
\, t^{\,\star} : \ = \  \frac{t}{t_{\, ref}} ; \; \; \;
\, x^{\,\star}  \ = \  \frac{x}{L}  ; \; \; \;
\, Fo : \ = \  \frac{ k_0 \ .  \, t_{ref} }{ c_{ 0} \, L^2}  \ = \ 1
\end{equation*}
\begin{equation*}
\, k^{\,\star}  \ = \  \frac{k}{k_0}  ; \; \; \;
\, c^{\,\star}  \ = \  \frac{c}{c_0}  ; \; \; \;
\, Bi_{\,in} : \ = \  \frac{h_{c, in}.L}{ k_0}  ; \; \; \;
\, Bi_{\,out} : \ = \  \frac{h_{c, out}.L}{ k_0} ; \; \; \;
\,  t_{\,ref}  : \ = \ \frac{c_0 \, L^2}{ k_0  }  ; \; \; \;
\end{equation*}
\begin{equation*}
\, T_{\, air, in}^{\,\star} : \ = \  \ - \ 1 \ + \  \frac{T_{\,air, in}}{T_{0}} ; \; \; \;
\, T_{\, air, out}^{\,\star} : \ = \  \ - \ 1 \ + \  \frac{T_{\,air, out}}{T_{0}} ; \; \; \;
\, T^{\,\star}_{\infty} : \ = \  \ - \ 1 \ + \  \frac{T_{\infty}}{T_{0}} ; \; \; \;
%\, \Gamma =  \frac{\tau}{t_{\, ref}} 
\end{equation*} 
\begin{equation*}
\, q_{\, net, in}^{\,\star}  : \ = \  \frac{q_{\, net, in} \, . \, L }{   k_0 \,  . \, T_{0} }  ; \; \; \;
\, q_{\, net, out}^{\,\star}  : \ = \  \frac{q_{\, net, out} \, . \, L }{   k_0 \,  . \, T_{0} }  ; \; \; \;
\end{equation*}
\nomenclature[B]{$Bi$}{Biot Number}
\nomenclature[B]{$Fo$}{Fourier Number}
\nomenclature[B]{$k$}{Thermal conductivity  $[ \mathsf{W}.\mathsf{m}^{-1}.\mathsf{K}^{-1} ]$  }
\nomenclature[B]{$c$}{Volumetric heat capacity  $[ \mathsf{J}.\mathsf{m}^{-3}.\mathsf{K}^{-1} ]$  }
\nomenclature[B]{$h_{c}$}{surface heat transfer coefficient  $[ \mathsf{W}.\mathsf{m}^{-2}.\mathsf{K}^{-1} ]$  }
\nomenclature[B]{$q_{net}$}{Net radiative heat flux  $[ \mathsf{W}.\mathsf{m}^{-2} ]$  }
\nomenclature[B]{$q_{SW}$}{Short-wave radiative heat flux  $[ \mathsf{W}.\mathsf{m}^{-2} ]$  }
\nomenclature[B]{$q_{LW}$}{Long-wave radiative heat flux  $[ \mathsf{W}.\mathsf{m}^{-2} ]$  }
\nomenclature[B]{$T_{air}$}{Air temperature  $[\mathsf{K} $ ] }
\nomenclature[B]{$T_{\infty}$}{Temperature in the ground [ $\mathsf{K} ]$  }
\nomenclature[B]{$L$}{Domain characteristic length  $[ \mathsf{m} ] $  }
\nomenclature[B]{$_{in/out}$}{Location: inside or outside  }
\bigbreak
\textbf{Inside air temperature}\\ 
\bigbreak
The problem can be rewritten in the following dimensionless form:  
\begin{align}
    \frac{ d  T^{\star}_{\, air, \, in} }{ d  \, t ^{\star}} (t) & \ = \ q_{v}^{\star} \ + \sum_{j \ = \ 1}^{\mathcal{P}} \theta_j \, .  \, \Bigl(\, T_{j}^{\star}  \ - \ T_{\,in}^{\star} \, \Bigr). 
\end{align}
with the following dimensionless numbers  :
\begin{align*}
\, q_{v}^{\,\star} : \ = \  \frac{q_{v}}{q_{v,0}} ; \; \; \;
\, \theta_j : \ = \  \frac{S_{\,j}. h_{c, in} . t_{0}}{  V_{air} \, c_{air} }  \; \; \; \, 
\end{align*}

\bibliographystyle{unsrt}
\bibliography{sample}

\end{document}